
\input epsf	



\magnification=\magstephalf

\def\dbar{{\overline d}}
\def\gam{{`$\Gamma$'}}
\def\sigm{{\sigma_{8,{\rm mass}}}}
\def\sigg{{\sigma_{8,{\rm gal}}}}
\def\rhom{{\rho_{{\rm mass}}}}

\def\rhomavg{{\overline{\rho}_{{\rm mass}}}}
\def\rhogavg{{\overline{\rho}_{{\rm gal}}}}
\def\ng{{n_{{\rm gal}}}}
\def\np{{n_{{\rm p}}}}
\def\ngavg{{\overline{n}_{{\rm gal}}}}

\newbox\grsign \setbox\grsign=\hbox{$>$} \newdimen\grdimen \grdimen=\ht\grsign
\newbox\simlessbox \newbox\simgreatbox
\setbox\simgreatbox=\hbox{\raise.5ex\hbox{$>$}\llap
     {\lower.5ex\hbox{$\sim$}}}\ht1=\grdimen\dp1=0pt
\setbox\simlessbox=\hbox{\raise.5ex\hbox{$<$}\llap
     {\lower.5ex\hbox{$\sim$}}}\ht2=\grdimen\dp2=0pt

\def\simlt{\mathrel{\copy\simlessbox}}

\def\textskip{\baselineskip = 12pt plus 1pt minus 1pt}
\def\capskip{\baselineskip = 10 pt}


\font\chpfont=cmbx10 scaled\magstep1

\def\b#1{\skew{-2}\vec#1}  
%

%


\def\spose#1{\hbox to 0pt{#1\hss}}

\font\eightrm=cmr8

\def\s{\ifmmode \widetilde \else \~\fi} 

\def\={\overline}
\def\section{\S}
\newcount\notenumber
\notenumber=1
\newcount\eqnumber
\eqnumber=1
\newcount\fignumber
\fignumber=1
\newbox\abstr


\def\hmpc{h^{-1}\/{\rm Mpc}}

\def\note#1{\footnote{$^{\the\notenumber}$}{#1}\global\advance\notenumber by 1}
\def\foot#1{\raise3pt\hbox{\eightrm \the\notenumber}
     \hfil\par\vskip3pt\hrule\vskip6pt
     \noindent\raise3pt\hbox{\eightrm \the\notenumber}
     #1\par\vskip6pt\hrule\vskip3pt\noindent\global\advance\notenumber by 1}

\def\abstract#1{\setbox\abstr=\vbox{\hsize 5.0truein{\par\noindent#1}}
    \centerline{ABSTRACT} \vskip12pt \hbox to \hsize{\hfill\box\abstr\hfill}}

\def\Dt{\spose{\raise 1.5ex\hbox{\hskip3pt$\mathchar"201$}}}	
\def\dt{\spose{\raise 1.0ex\hbox{\hskip2pt$\mathchar"201$}}}	

\def\new{{\rm\chaphead\the\eqnumber}\global\advance\eqnumber by 1}
\def\ref#1{\advance\eqnumber by -#1 \chaphead\the\eqnumber
     \advance\eqnumber by #1 }
\def\last{\advance\eqnumber by -1 {\rm\chaphead\the\eqnumber}\advance 
     \eqnumber by 1}
\def\eqnam#1{\xdef#1{\chaphead\the\eqnumber}}

\def\refindent{\par\noindent\parskip=4pt\hangindent=3pc\hangafter=1 }

\def\refbook#1{\refindent#1}

\def\sectionbegin#1{\vskip15pt\par\noindent{\bf#1}\par\vskip8pt}
\def\subsectionbegin#1{\vskip8pt\par\noindent{\bf#1}\par\vskip4pt}
\def\subsubsectionbegin#1{\vskip8pt\par\noindent{\bf#1}\par\vskip4pt}

\def\lta{\mathrel{\spose{\lower 3pt\hbox{$\mathchar"218$}}
     \raise 2.0pt\hbox{$\mathchar"13C$}}}
\def\gta{\mathrel{\spose{\lower 3pt\hbox{$\mathchar"218$}}
     \raise 2.0pt\hbox{$\mathchar"13E$}}}


\def\ie{\hbox{\it i.e.,\ }}
\def\eg{\hbox{\it e.g.,\ }}
\def\etal{\hbox{\it et al.\ }}


\def\ccom{\,\raise2pt\hbox{,}} 

\def\dn#1{\lower 2 pt\hbox{$\scriptstyle #1$}}
\def\up#1#2{\raise {#1} pt\hbox{$\scriptstyle #2$}}
\def\hprime{\raise 2pt\hbox{$\scriptstyle \prime$}}



\def\sectionskip{\penalty-500\vskip24pt plus12pt minus6pt}

\def\SECTION#1#2\par{\goodbreak
   \sectionskip\leftline{\bf #1\quad #2}\nobreak\vskip0.3cm
   \mark{#1\quad #2}} 
     \parskip  = 6 true  pt plus 1 true pt minus 1 true  pt


\font\ninerm=cmr9   \font\eightrm=cmr8   \font\sixrm=cmr6
\font\ninei=cmmi9   \font\eighti=cmmi8   \font\sixi=cmmi6
\font\ninesy=cmsy9  \font\eightsy=cmsy8  \font\sixsy=cmsy6
\font\ninebf=cmbx9  \font\eightbf=cmbx8  \font\sixbf=cmbx6
\font\ninett=cmtt9  \font\eighttt=cmtt8  
\font\nineit=cmti9  \font\eightit=cmti8  
\font\ninesl=cmsl9  \font\eightsl=cmsl8

\catcode`@=11
\newskip\ttglue


\def\tenpoint{\def\rm{\fam0\tenrm}
  \textfont0=\tenrm \scriptfont0=\sevenrm \scriptscriptfont0=\fiverm
  \textfont1=\teni  \scriptfont1=\seveni  \scriptscriptfont1=\fivei
  \textfont2=\tensy \scriptfont2=\sevensy \scriptscriptfont2=\fivesy
  \textfont3=\tenex \scriptfont3=\tenex \scriptscriptfont3=\tenex
  \textfont\itfam=\tenit  \def\it{\fam\itfam\tenit}%
  \textfont\slfam=\tensl  \def\sl{\fam\slfam\tensl}%
  \textfont\ttfam=\tentt  \def\tt{\fam\ttfam\tentt}%
  \textfont\bffam=\tenbf  \scriptfont\bffam=\sevenbf
   \scriptscriptfont\bffam=\fivebf  \def\bf{\fam\bffam\tenbf}%
  \tt \ttglue=0.5em plus .25em minus.15em
  \setbox\strutbox=\hbox{\vrule height8.5pt depth3.5pt width0pt}%
  \let\sc=\eightrm  \let\big=\tenbig  \normalbaselines\rm}

\def\ninepoint{\def\rm{\fam0\ninerm}
  \textfont0=\ninerm \scriptfont0=\sixrm \scriptscriptfont0=\fiverm
  \textfont1=\ninei  \scriptfont1=\sixi  \scriptscriptfont1=\fivei
  \textfont2=\ninesy \scriptfont2=\sixsy \scriptscriptfont2=\fivesy
  \textfont3=\tenex  \scriptfont3=\tenex \scriptscriptfont3=\tenex
  \textfont\itfam=\nineit  \def\it{\fam\itfam\nineit}%
  \textfont\slfam=\ninesl  \def\sl{\fam\slfam\ninesl}%
  \textfont\ttfam=\ninett  \def\tt{\fam\ttfam\ninett}%
  \textfont\bffam=\ninebf  \scriptfont\bffam\sixbf
   \scriptscriptfont\bffam=\fivebf  \def\bf{\fam\bffam\ninebf}%
  \tt \ttglue=0.5em plus .25em minus.15em
  \setbox\strutbox=\hbox{\vrule height8pt depth3pt width0pt}%
  \let\sc=\sevenrm  \let\big=\ninebig  \normalbaselines\rm}

\def\eightpoint{\def\rm{\fam0\eightrm}
  \textfont0=\eightrm \scriptfont0=\sixrm \scriptscriptfont0=\fiverm
  \textfont1=\eighti  \scriptfont1=\sixi  \scriptscriptfont1=\fivei
  \textfont2=\eightsy \scriptfont2=\sixsy \scriptscriptfont2=\fivesy
  \textfont3=\tenex \scriptfont3=\tenex \scriptscriptfont3=\tenex
  \textfont\itfam=\eightit  \def\it{\fam\itfam\eightit}%
  \textfont\slfam=\eightsl  \def\sl{\fam\slfam\eightsl}%
  \textfont\ttfam=\eighttt  \def\tt{\fam\ttfam\eighttt}%
  \textfont\bffam=\eightbf  \scriptfont\bffam\sixbf 
   \scriptscriptfont\bffam=\fivebf \def\bf{\fam\bffam\eightbf}%
  \tt \ttglue=0.5em plus .25em minus.15em
  \setbox\strutbox=\hbox{\vrule height7pt depth2pt width0pt}%
  \let\sc=\sixrm  \let\big=\eightbig  \normalbaselines\rm}
\def\tenbig#1{{\hbox{$\left#1\vbox to8.5pt{}\right.\n@space$}}}
\def\ninebig#1{{\hbox{$\textfont0=\tenrm\tenfont2=\tensy
  \left#1\vbox to7.25pt{}\right.\n@space$}}}
\def\eightbig#1{{\hbox{$\textfont0=\ninerm\tenfont2=\ninesy
  \left#1\vbox to6.5pt{}\right.\n@space$}}}



%

         \newtoks\chaptitle
         \newcount\startpage

\def\b#1{\skew{-2}\vec#1}

\def\abstract#1{\setbox\abstr=\vbox{\hsize 5.0truein{\par\noindent#1}}
    \centerline{ABSTRACT} \vskip12pt \hbox to \hsize{\hfill\box\abstr\hfill}}

\def\sectionbegin#1{\vskip15pt\par\noindent{\bf#1}\par\vskip8pt}
\def\subsectionbegin#1{\vskip10pt\par\noindent{\bf#1}\par\vskip6pt}

\def\lta{\mathrel{\spose{\lower 3pt\hbox{$\mathchar"218$}}
     \raise 2.0pt\hbox{$\mathchar"13C$}}}
\def\gta{\mathrel{\spose{\lower 3pt\hbox{$\mathchar"218$}}
     \raise 2.0pt\hbox{$\mathchar"13E$}}}



\centerline{\chpfont Cosmic Voids and Biased Galaxy Formation}
\vskip 0.1 truecm
\centerline{\bf Blane Little$^{1}$ and David H. Weinberg$^{2}$} 
\vskip 0.3 truecm

\line{$^1$ {\it Astronomy Unit, School of Mathematical Sciences,
Queen Mary and Westfield College,}\hfil}
\line{\ \ \ \ {\it Mile End Road, London E1 4NS}\hfil} 

\line{$^2$ {\it 
Institute for Advanced Study, 
Olden Lane,
Princeton, New Jersey, 08540 U.S.A.}\hfil} 

\vskip 0.3truecm
\line{e-mail: bl@starlink.qmw.ac.uk, dhw@guinness.ias.edu \hfil}

\vskip 0.5truecm
\textskip

\sectionbegin {\bf Abstract} 

Using cosmological $N$-body simulations and the void probability
function (VPF), we investigate the statistical properties of 
voids within a wide range of initially Gaussian models for the origin of
large-scale structure. We study the dependence of the VPF on
cosmological parameters, on the power spectrum of primordial
fluctuations, and on assumptions about galaxy formation. We pay
particular attention to the ability of the VPF to diagnose `biased
galaxy formation': the preferential formation of galaxies in regions of
high background density and corresponding suppression of galaxy formation
in regions of low background density.  
We find that the VPF is insensitive to the cosmic density parameter 
$\Omega_0$ and the cosmological constant $\lambda_0$, provided that
fluctuations are normalized to a fixed rms amplitude on scales 
$\sim 8$ $h^{-1}$ Mpc.  
In the absence of biasing, the VPF is also insensitive
to the shape of the initial power spectrum.
The VPF does depend on the prescription adopted for biased galaxy formation,
in the obvious sense that a scheme that more efficiently suppresses galaxy
formation in low density regions leads to larger voids.   Biased models
have systematically higher VPFs than unbiased models, but   for a given
biasing scheme the VPF is relatively insensitive to the value   of the
bias factor $b$, the ratio of rms galaxy fluctuations to rms   mass
fluctuations.  Thus, while the VPF can distinguish unbiased models  
from some biased models, it is probably not a useful way to constrain  
the bias factor; uncertainties in the appropriate choice of biasing  
prescription overwhelm the mild dependence on $b$. 

We compare the predictions of our models to the most extensive VPF 
observations published to date.  These data do {\it not\/} require strong
biasing; Gaussian models in which galaxies trace mass can reproduce
the VPF data to within the errors expected from the current finite
volume fluctuations.  Models with the moderate biasing predicted by
cosmological simulations that incorporate gas dynamics 
yield a slightly better match to the data.
Models in which galaxy formation is strongly
suppressed in low density regions produce an excess of large, empty voids.

\vfill\eject

\sectionbegin{1. Introduction} 

Giant voids are among the most striking features of the observed
distribution of galaxies ({\it e.g.\/} Gregory \& Thompson 1978;
Kirshner \etal 1981, 1987; Davis \etal 1982;
de Lapparent, Geller \& Huchra 1986; see review by Rood 1988 and
references therein).
The remarkable interlocking pattern of
superclusters and voids revealed by galaxy redshift surveys has
prompted various authors to describe the observed galaxy distribution 
as a ``cell structure'' (Joeveer \& Einasto 1978), a foam
of ``bubbles'' (de Lapparent, Geller \& Huchra 1986), or a ``sponge-like''
network of interlocking filaments and tunnels (Gott, Melott \& Dickinson 1986).
The most widely explored models for the origin of large-scale structure
propose that the clusters, superclusters, and voids that we observe
today developed by gravitational instability from small-amplitude, Gaussian
fluctuations, generated by physical processes in the very early universe.
Can the gravitational growth of Gaussian primordial fluctuations account for 
the observed voids, or do these models require that galaxy formation be
suppressed in low density regions in order to produce voids as large and
empty as observed?  Do voids represent regions where there is no
mass, or merely regions where there are no (bright) galaxies?  
In this paper we
address these questions using cosmological $N$-body simulations and the
void probability function (VPF), a simple statistical measure of the sizes
of voids.  The VPF of a galaxy 
sample is the probability $P_0(R)$ that a randomly
placed sphere of radius $R$ contains no galaxies.
We study the dependence of the VPF on cosmological parameters,
on the power spectrum of the primordial fluctuations, and -- above all 
--
on assumptions about galaxy formation.  We also compare our results to the
most extensive VPF observations published to date
(Vogeley, Geller \& Huchra 1991).

The nature of voids is intimately connected to the issue of
`biased galaxy formation,' an idea that first gained popularity in the context
of the cold dark matter (CDM) model of structure formation.
The most theoretically attractive version of CDM assumes a critical
density ($\Omega=1$) universe.  Observations of cluster mass-to-light
ratios, cluster velocity dispersions, and the galaxy pairwise velocity
dispersion clearly contradict this assumption {\it if} galaxies are
clustered in the same way as mass ({\it e.g.\/},
Davis {\it et al.\/} 1985, hereafter DEFW).  However, if galaxy formation
is more efficient (per unit mass) in regions of high background density,
then galaxies will cluster more strongly than the underlying mass distribution,
and an $\Omega=1$ model can, perhaps, be reconciled with the observations 
(DEFW; Bardeen {\it et al.\/} 1986, hereafter BBKS).  
While the term `biasing' might be used to describe any difference between the
large-scale galaxy and mass distributions, we will use it in the specific
sense mentioned above: preferential formation of galaxies in regions of
high background density.  If galaxies form more efficiently than average
in high density regions, then they must form less efficiently than average
in low density regions, so biasing naturally produces voids that are empty
of galaxies but not completely empty of mass.

There are observational and theoretical reasons for thinking that
biasing might be an important phenomenon, independent of the $\Omega=1$
assumption.  On the observational side, the well known morphology-density
relation implies that elliptical and spiral galaxies have different 
clustering properties (Dressler 1980; Postman \& Geller 1984).
At least one of these classes of galaxies must
cluster differently than the mass, and there is no particular reason to
think that either class individually or the union of the two traces
the large-scale mass distribution.  On the theoretical side,
galaxy-scale perturbations collapse earlier in regions of high background 
density, so they 
tend to reach higher internal densities and cool more efficiently than
equivalent perturbations in low density regions.
Indeed, numerical simulations that include gas dynamics indicate that galaxy
formation is at least somewhat biased towards regions of high background
density (Cen \& Ostriker 1992; 
Katz, Hernquist \& Weinberg 1992; 
see also
White \etal 1987; 
Kaiser 1988; 
Gelb 1992).

It has also been argued that gravity alone cannot create voids as large and
empty as those observed, and that the existence of these voids is itself
evidence for biased galaxy formation ({\it e.g.\/} Dekel \& Rees 1987;
Betancort-Rijo 1990;
Einasto {\it et al.\/} 1991, hereafter EEGS).  This line of reasoning, if
correct, enhances the plausibility of $\Omega=1$ models by providing
independent evidence for biasing, and it suggests that a measure of 
void sizes like the VPF might offer a sensitive statistical diagnostic
of biasing.  We will assess the strength of the VPF as a biasing diagnostic,
and we will examine the ability of current theoretical models to explain
the observed spectrum of void sizes.

There have been several recent $N$-body studies of the gravitational growth
and interactions of voids, in idealized configurations (Dubinski \etal 1993;
see also West, Weinberg \& Dekel 1990), in void-dominated structure models
(Reg\"os \& Geller 1991),
and in models with Gaussian initial conditions
(van de Weygaert \& van Kampen 1993).  
These papers did not examine the void probability function or the 
effects of biased galaxy formation, which will be the central concerns
of our investigation.
The two most direct predecessors of the present study are the above cited
work by EEGS and the paper by Weinberg \& Cole (1992; hereafter WC). The 
latter
applied the VPF (and other clustering statistics) to $N$-body models
with Gaussian and non-Gaussian initial conditions.
WC found the VPF to be a sensitive discriminant between Gaussian and 
non-Gaussian models in the absence of biasing, but they found, unsurprisingly,
that biasing could create large voids in all models.  Gaussian models
proved more successful than any of WC's non-Gaussian models in explaining
the full range of galaxy clustering data, and the Gaussian hypothesis has
received further observational support from recent analyses of large-scale
galaxy counts (Bouchet \etal 1993 and references therein).  
Since Gaussian fluctuations have both theoretical simplicity and a degree 
of empirical success on their side, we will restrict our attention to 
Gaussian models in this paper.

Because of our relatively narrow focus, we have been able to improve
upon the studies of EEGS and/or WC in a number of ways: 

\item {(1)} EEGS's models were normalized by arranging for them to have
the same final rms mass fluctuation on a particular scale.
However, the distribution of total (luminous + dark) mass
in the real universe is not
well-determined.
A more meaningful comparison between theory and
observation can therefore 
be obtained if one's theoretical models are adjusted to
share an {\it observable\/} property, such as the rms fluctuation in galaxy
counts on a particular scale. 
This is what we do, and it constitutes the most important
difference of principle between our study and that of EEGS. 

\item {(2)} The initial conditions of $N$-body simulations cannot
include contributions from Fourier density waves that are bigger than
the simulation cube, and the enforcement of periodic boundary conditions 
during $N$-body evolution prevents such power from developing.
A comparable volume of the real universe does not suffer from these restraints.
$N$-body simulations systematically underestimate the frequency of large
voids if the simulation volume is not large enough to contain all waves that
significantly influence the growth of underdense regions.
This limitation may have
affected the results of EEGS and WC, whose periodic 
$N$-body cubes had sides of 40 
and 128 $h^{-1}$ Mpc, respectively
($h\equiv H_0 / 100$ km s$^{-1}$ Mpc$^{-1})$. By contrast, Kauffmann \& Melott
(1992) suggest that a CDM simulation
would require a $\sim 160$ $h^{-1}$ Mpc box `before one could begin to
trust the typical size of voids'.
We use simulation cubes that are significantly
larger than those of EEGS and WC -- either $300$ or 192 $h^{-1}$ Mpc on a side.
The large simulation volume minimizes the systematic effects of missing
large-scale waves, and it reduces the statistical uncertainty
in our estimates of the VPF. 

\item {(3)} EEGS and WC employ a single biasing prescription, which we 
shall call `density biasing'.  Both sets of authors
evolve an ensemble of mass points to redshift $z=0$, smooth the
resultant density field, and identify as `galaxies' only those particles
above some sharp cutoff in local density. We also investigate this biasing
prescription, but we examine two
other biasing schemes as well. One of these is the well-known `peaks
biasing' prescription, which identifies galaxies with high peaks of the initial
density field (\eg DEFW; BBKS).  For our simulations of the standard CDM
model, we also consider the biasing relation derived by Cen \& Ostriker (1993)
from their hydrodynamic simulations of this scenario.
Our three biasing prescriptions differ substantially, and together 
they span a range of physically relevant possibilities.

\item {(4)} EEGS examined biased versions of only 3 broad types of
theoretical models: a distribution of randomly placed galaxies at $z=0$;
a distribution of randomly placed clusters at $z=0$; and a
numerically evolved, spatially flat, Gaussian model with the 
initial power spectrum of $\Omega _0 =0.2$ CDM.
Only the last of these exhibits the complex clustering characteristic
of the observed galaxy distribution.
WC evolved 32 theoretical models, but three quarters of them
were associated with non-Gaussian initial conditions, and they used only
pure power-law initial spectra.
We will present some results for power-law spectra, but will concentrate
our analysis on initial 
power spectra with stronger theoretical and empirical
motivation.  We do not consider non-Gaussian models, but we achieve much
better coverage of the most physically promising cosmological 
models than either WC or EEGS.

\item {(5)}  The VPF depends rather sensitively on the number
density $n$ of the galaxy sample used to measure it --
the sparser the
sample, the bigger the voids. 
While EEGS computed only the VPF, we also compute a related statistic --
the underdense probability function (`UPF' or $P_{80}$) -- that is
independent of $n$ (except for shot noise, which becomes unimportant at
large void radii). 

\item {(6)} WC examined only two biasing strengths: $b=1$ (unbiased) and
$b=2$, where $b$ is the `bias factor' defined by the relation
$\sigma _{8,{\rm galaxies}} = b\;\sigma _{8,{\rm mass}}.$  Here $\sigma_8$ is
the rms fluctuation of the density contrast in randomly placed spheres
of radius 8 $h^{-1}$ Mpc.  While we also 
explore $b=1$ and $b=2$, we investigate two additional biasing
strengths of interest ($b=1.5$ and $b=3$). 

\item {(7)} We improve upon the comparison between theory and
observation in three respects: 

\itemitem {({\it i\/})} The VPF data are in redshift space, but EEGS and
WC compare them with theoretical predictions computed in real
space. This can be misleading if the move from real space to redshift
space significantly changes the VPF. We compute the VPF both in real
space and in redshift space, but -- as consistency requires -- we
compare the VPF data only with redshift-space predictions. 

\itemitem {({\it ii\/})} 
Vogeley, Geller, \& Huchra (1991, hereafter VGH)
present
the most extensive VPF data published to date.
Most of these data were unavailable to EEGS, and WC
use results
from only 2 of VGH's 12 volume-limited samples. We compare our
predictions with all 12 of these samples, some of which extend to 
126
$h^{-1}$ Mpc.
The samples used by WC and
EEGS all had limiting depths of $\lta 80$ $h^{-1}$ Mpc. 

\itemitem {({\it iii\/})} Unlike both EEGS and WC,
we quantify the effects of finite volume errors on current estimates of the 
VPF.

The next section provides a detailed description of our models, \S 3
presents our main results, and \S 4 summarizes our conclusions. 

\vfill\eject

\sectionbegin{2. The Models} 

\sectionbegin{2.1 Initial Conditions} 

Our basic set of initial conditions consists of an isotropic Gaussian
random field (GRF) of density fluctuations,
$\delta (\b r) \equiv [\rho (\b r) -\overline\rho] / \overline\rho$,
defined within a triply-periodic simulation cube.
Such a field can be expanded in terms of complex Fourier components
$\delta _{\b k} = A_{\b k}\;e^{i\theta _{\b k}}$,
and its statistical properties are completely 
determined by its power spectrum
$P(k)\equiv \langle |\delta_{\b k}|^2\rangle$.
For a GRF, the phases $\theta _{\b k}$ are 
randomly distributed in the interval [0, $2\pi$], and
the amplitudes $A_{\b k}$ are Rayleigh-distributed with variance
$P(k)$. We ensure these 
conditions by assigning values
drawn from a Gaussian distribution with variance $P(k)/2$ separately to 
the real and imaginary parts of each $\delta _{\b k}$
(cf. BBKS). These Fourier components then allow the computation of a 
GRF $\delta (\b r)$
with a specified power spectrum $P(k)$.

The three cosmological scenarios that we examine in greatest detail have 
initial power spectra described by equation (7) of 
Efstathiou, Bond, \& White (1992; hereafter EBW), with two different
values of their `$\Gamma$' parameter.
With $\Gamma=\Omega_0 h$, this equation provides an accurate fit to the
post-recombination, linear power spectrum of adiabatic fluctuations 
that enter the horizon with scale-invariant amplitudes and grow thereafter
in a universe dominated by cold dark matter.
Our first model is `standard CDM', with $\Omega=1$ and $h=0.5$
(implying $\Gamma=0.5$).  While the standard CDM model has many 
attractive features, a number of observations suggest that the
$\Gamma=0.5$ spectrum lacks
sufficient power on large scales (\eg 
Efstathiou \etal 1990; 
Maddox {\it et al\/.} 1990; 
Saunders \etal 1991; Moore \etal 1992;
Vogeley \etal 1992; Fisher \etal 1993).
A better fit to galaxy clustering data is obtained for $\Gamma=0.25$,
and we consider two models with this spectrum.  The first is an
$\Omega_0=0.3$ CDM model, which naturally produces this power spectrum
if $h \approx 0.8.$  For our standard version of this model we include
a cosmological constant, $\lambda_0 \equiv \Lambda /3H_0^2=0.7$,
so that the spatial curvature vanishes,
as predicted by inflationary cosmology
(see review by Narlikar \& Padmanabhan 1991).
We also examine an $\Omega=1$ model with $\Gamma=0.25$.  There are
various physical scenarios that could produce a $\Gamma \approx 0.25$
spectrum in an $\Omega=1$ universe.  They include models in which
equalization occurs late because a decaying particle enhances the background
neutrino density (\eg Bond \& Efstathiou 1991) or because the Hubble constant
is very low.  One particularly interesting possibility
is a universe dominated by a mixture of hot and cold dark matter --
such a model may best explain 
galaxy clustering and velocity statistics, as well as
the microwave background fluctuations recently measured by COBE
(\eg EBW; 
Davis, Summers \& Schlegel 1992; 
Taylor \& Rowan-Robinson 1992; 
Klypin \etal 1993). 
From a theoretical perspective, all of the proposed models that yield
a $\Gamma=0.25$ spectrum seem somewhat contrived, but one can remain
agnostic on the subject of theoretical underpinning and simply view
the $\Gamma=0.25$ spectrum as a reasonably successful empirical fit to
observational data.

Figure 1 shows the $\Gamma=0.5$ and $\Gamma=0.25$ initial power spectra,
extrapolated via linear perturbation theory to $z=0$ and normalized so that 
$\sigm=1$, \ie so that the rms mass fluctuation is unity
in randomly placed spheres of radius 8 $h^{-1}$ Mpc.  
The $\Gamma=0.25$ spectrum has more power on scales larger
than this normalization scale.
The vertical arrows in this plot indicate the lowest and highest
comoving wavenumbers included in our initial conditions: $k/k_{{\rm f}}=1$ (the 
wavenumber of the fundamental mode of our 300 $h^{-1}$ Mpc simulation box)
and $k/k_{{\rm f}}=50$ (the Nyquist frequency of the $100^3$ grid used 
to set up our initial conditions). 
Note that both of the spectra in Figure 1 approach an $n=1$ power law as
$k\rightarrow 0$, in agreement both with 
inflationary cosmology (see Narlikar \& Padmanabhan 1991) 
and with COBE's recent 
observations of temperature fluctuations in the cosmic microwave background
radiation (Smoot {\it et al.\/} 1992). 

\topinsert
\capskip
\centerline{
\epsfysize=3.0truein
\epsfbox[72 396 540 738]{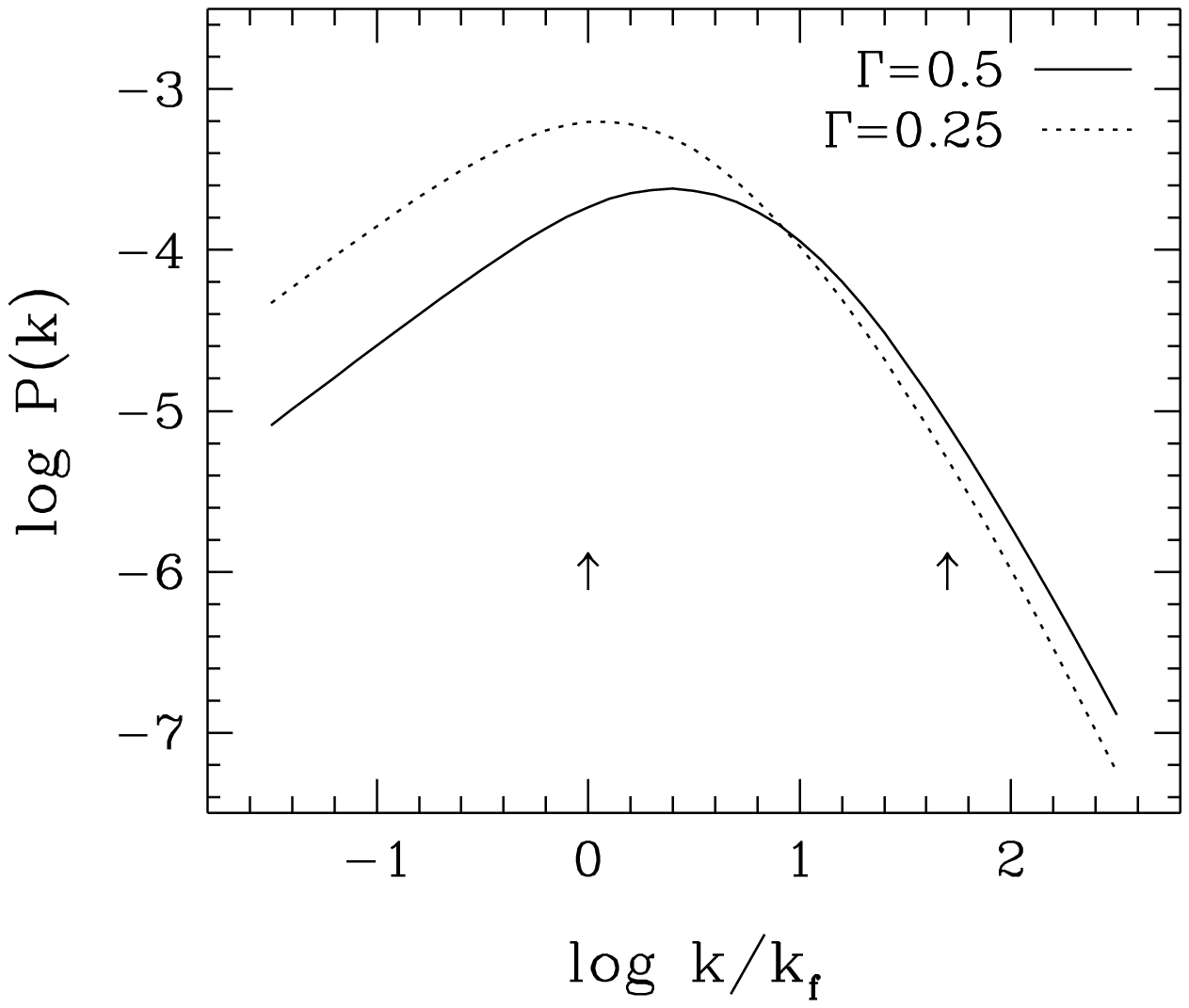}
}
\medskip
{\eightpoint
\capskip
\noindent {\bf Figure 1} --- 
Initial power spectra for our $\Gamma$ = 0.5 and $\Gamma$ = 0.25 models.
Both spectra have been extrapolated via linear perturbation theory to
$z=0$, and normalized so that the rms mass fluctuation is unity in
randomly placed spheres of radius 8 $h^{-1}$ Mpc. On scales larger than
this, the $\Gamma$ = 0.25 spectrum has more power than the $\Gamma$ =
0.5 spectrum.  Vertical arrows indicate the lowest and highest
comoving wavenumbers included in our initial conditions:
$k=k_{\rm f}$ (the wavenumber of the fundamental mode of our
300 $h^{-1}$ Mpc simulation box) and $k$ = 50 $k_{\rm f}$ (the Nyquist
frequency of the $100^3$ grid used to set up our initial conditions). 
}
\medskip
\textskip
\endinsert

In order to study the dependence of the VPF on cosmological parameters,
we also consider models with the $\Gamma=0.25$ power spectrum and
$(\Omega_0,\lambda_0)$ = (0.3, 0), (0.1, 0.9), and (0.1, 0).
We do not change the power spectrum in concert with $\Omega_0$ because
we want to separate dependence on cosmological parameters from dependence
on the power spectrum.  We will show in \S 3.3 that the VPF is
insensitive to $\Omega_0$ and $\lambda_0$ if the power spectrum,
normalization, and biasing scheme are held fixed.  We therefore do not
carry out a detailed comparison between these last three models
and observational data.

The $\Gamma=0.25$ and $\Gamma=0.5$ power spectra are not radically
different over the range of scales probed by our simulations.
We want to study the dependence of the VPF on the shape of the initial
spectrum using a somewhat larger lever arm than these models provide,
so we also examine models with pure power-law initial spectra,
$P(k) \propto k^n$.  A pure power-law spectrum that failed to turn over
or cut off would violate observational constraints on either very
large or very small scales, but a power law might be a reasonable 
approximation to the true power spectrum over some finite range.
`Standard CDM' predicts a post-recombination power spectrum 
approximately like $P(k) \propto k^{-1}$
on comoving scales $\sim 3-10$ $h^{-1}$ Mpc, and a 
variety of observations support this prediction
(Gott \& Rees 1975; Gott \& Turner 1977; Bhavsar, Gott \& Aarseth
1981; Gunn 1982; Gott {\it et al.\/} 1989, 1992). 
For our power-law spectra, we therefore consider
$n=-1$ and its `bracketing values' of $n=-2$ and $n=0$. 

\vfill\eject
\sectionbegin{2.2 Non-linear evolution} 

We evolve our initial conditions into the non-linear regime using a
particle-mesh (PM) $N$-body code written by Changbom Park. This code
employs a staggered-mesh technique (Melott 1986) to achieve higher force
resolution than a conventional PM code. The code is thoroughly described
by Park (1990), and it has been tested against analytic solutions and
other $N$-body codes (Park 1990; Weinberg {\it et al.\/}, in
preparation, hereafter W+). The W+ tests show that this PM code
reproduces the results of high-resolution P$^3$M (\eg Efstathiou {\it et
al.\/} 1985) and tree (\eg Hernquist, Bouchet, \& Suto 1991) $N$-body
codes down to the limit of its force resolution, about 1-2 mesh cells.
The chief advantage of the PM method is speed -- P$^3$M or tree code
simulations with the same mass resolution as our PM experiments would
have taken several times longer to perform. 

PM simulations require a trade-off between force resolution and the size
of the simulation volume, since the dimensions of the computational grid
are usually limited by memory and cpu constraints.
We need a large simulation volume for the reasons discussed in \S 1,
but because the VPF is insensitive to small-scale clustering we can get away
with relatively low force resolution.
Our simulations of `$\Gamma$' models (those with a CDM-like power spectrum)
use $100^3$ particles on a $200^3$ density-potential mesh to represent
a $300$ $h^{-1}$ Mpc, comoving, periodic cube.  For simulations with 
power-law initial 
spectra (about whose realism we are somewhat less concerned),
we use $64^3$ particles on a $128^3$ mesh, and a $192$ $h^{-1}$ Mpc simulation
cube.  In all cases the mesh scale is $\sim 1.5$ $h^{-1}$ Mpc.
We have used trial runs of smaller volumes to confirm that we obtain the
same VPF with a force resolution (mesh scale) of 0.5, 0.75, 1.0, and 
1.5 $h^{-1}$ Mpc.  
We have also checked that the VPF is unaffected by
using 1 particle per 8 cells during dynamical evolution
instead of 1 particle per cell;
the sparser particle grid 
allows us to use a larger simulation volume with a fixed computer memory.
Although
we use a $200^3$ mesh (or $128^3$ for power-law models) to compute
forces during non-linear evolution, we set up our initial conditions using
a $100^3$ (or $64^3$) density field so that we do not alias high frequency
power into our initial particle distributions.

We use the Zel'dovich
approximation to turn our initial density fields into initial positions
and growing-mode velocities for our $N$ particles 
(see Doroshkevich \etal 1980; Efstathiou \etal 1985).
All our simulations
begin at a redshift $z_{\rm init}$ and advance to the present epoch
in $z_{\rm init}$ timesteps, which are equally spaced in the
expansion factor $a(t)$. 
We use $z_{\rm init}=31$ for the `$\Gamma$' model simulations, and
$z_{\rm init}=24$ for the power-law simulations.  Again, we have used trial
runs on smaller volumes to check our choices of initial redshift and timestep.
Doubling $z_{\rm init}$ or the number of timesteps would not alter the VPF
of our simulations.  Halving $z_{\rm init}$ or the number of timesteps would
have a tiny impact in a few cases.  We computed two realizations of each
of the `$\Gamma$' models, and four realizations of each of the power-law models.

\sectionbegin{2.3 Normalization}

The amplitude of primordial density fluctuations cannot presently
be calculated {\it a priori\/} from theory, and we treat it as a 
free parameter of our models.  The choice of amplitude is important,
since larger amplitude perturbations produce stronger clustering and
larger voids.  [In an $\Omega=1$ model, choosing the amplitude normalization
is equivalent to choosing the time in an evolving $N$-body simulation that
corresponds to the present, and voids grow as a simulation evolves.]
We normalize our models by making them fit an observational constraint, 
{\it viz.\/}, the rms
fluctuation of galaxy counts in spheres of radius 8 $h^{-1}$ Mpc
at the present epoch $t_0$.
Estimates of the galaxy correlation function (\eg Davis \& Peebles 1983)
imply that this quantity is close to unity, \ie $\sigma _{8,{\rm
gal}}(t_0) \simeq 1$.
This condition and the relation
$\sigma _{8,{\rm gal}} = b\;\sigma _{8,{\rm mass}}$ imply
that for our unbiased models -- where galaxies are assumed to trace the
mass distribution ($b=1$) -- the present epoch can be identified as
the time when the rms mass fluctuation in spheres of radius 8 $h^{-1}$
Mpc equals one, \ie $\sigma _{8,{\rm mass}}\simeq 1$. 

In the \gam\ models, the quantity $\sigm$ evolves at close to its linear
theory rate throughout our simulations.  We therefore normalize our initial
conditions so that they have $\sigma_8=1$ if extrapolated to $z=0$ by 
linear theory.
For some of the power-law models, particularly $n=0$, the growth of 
$\sigma_8$ departs significantly from linear theory, as discussed by WC.  
We adopt the normalization amplitudes computed by WC (see their table 1), so 
that the non-linear values of $\sigma_8$ in the simulations are equal to one 
at $z=0$.  The corresponding linear theory amplitudes are $\sigma_8=1.48$,
1.03, and 0.95 for $n=0$, $-1$, and $-2$, respectively.

The VPF is a sensitive function of the mean galaxy density $n$, or
equivalently, of the characteristic inter-galaxy separation $\overline
d\equiv n^{-1/3}$ (increasing the separation of galaxies 
increases the sizes of
voids). It is therefore important that our simulated data have
the same density
as the real data to which they are being compared. Once we
evolve our normalized initial conditions to $z=0$, we therefore
require that our simulated galaxies
have the characteristic 
separation
$\overline d = 4.5$ $h^{-1}$ Mpc
of VGH's densest sample.
To create an unbiased galaxy
population with 
$\overline d = 4.5$ $h^{-1}$ Mpc, we randomly
sample the particles in our final mass distribution to this density.

The normalization scheme discussed so far ignores the possibility of
biased galaxy formation. 
To create biased simulations with bias factors $b=1.5$, 2, and 3,
we reduce the amplitudes of the initial mass fluctuations by a factor of 
$b$,
and evolve them to $z=0$ by an $N$-body simulation as before.  We then
select a biased (instead of random) subset of the particles to represent
galaxies.  Our density biasing and peaks biasing prescriptions each contain
an adjustable parameter that controls the strength of clustering in the
biased particle subset.  We select this parameter so that $\sigg=1$.
To the extent that $\sigm$ follows linear theory, the ratio of rms galaxy
fluctuations to rms mass fluctuations at the final epoch
on the scale of 
8 $h^{-1}$ Mpc is $b$.

Instead of trying a variety of bias factors, we could have normalized the
mass fluctuations in our models by matching the amplitude of 
temperature fluctuations in the cosmic microwave background (CMB),
as observed by COBE (Smoot {\it et al.\/} 1992).  
We did not take this approach, for several
reasons.  First, predicting CMB fluctuations on the large angular scales
probed by COBE would require us to extrapolate our model power spectra to
wavelengths much larger than those that influence the formation of voids.
This extrapolation would make little physical sense for power-law spectra,
and even for the \gam\ models the resulting normalization would be unduly
sensitive to our precise choice of parameters.  For example, lowering
$\Gamma$ from 0.25 to 0.2 at fixed $\sigma_8$ would have a negligible
impact on void sizes, but it would significantly alter the relation between
$\sigma_8$ and COBE-scale CMB fluctuations.
Second, there are still significant random and systematic uncertainties
in the COBE fluctuation amplitude.  Finally, it is possible that some
fraction of the COBE signal arises from tensor mode (gravity wave) fluctuations
instead of the scalar mode fluctuations that correspond to density 
perturbations (Davis \etal 1992; Liddle \& Lyth 1992; Lidsey \& Coles 1992;
Lucchin \etal 1992; Salopek 1992; Souradeep \& Sahni 1992).
For what it is worth, EBW find that the best-fit COBE amplitude implies
$\sigma_8 \approx $ 1.1, 0.5, and 1.25 respectively for our spatially flat
models with $(\Gamma,\Omega_0)=$ (0.5, 1), (0.25, 1), and (0.25, 0.3),
assuming a truly scale-invariant primeval spectrum and no contribution from
gravity waves.

\goodbreak
\sectionbegin{2.4 Biasing Schemes}

In principle, the existence and nature of biased galaxy formation should be 
a prediction of a theoretical model, not an input.  A complete theory
specifies both the initial conditions and the important physical processes
for structure formation, and from these one should be able to compute
where the galaxies form and how they cluster relative to the mass.
In practice, the necessary computations are very difficult, and there
are uncertainties in the appropriate treatment of gas physics and star 
formation.  Most large-scale structure simulations therefore rely
on simplified prescriptions for identifying ``galaxy'' particles.
Simplest of all is the assumption that galaxies trace the mass,
which is the rule we adopt in our unbiased models.
We also consider two different prescriptions for biased galaxy formation,
one that identifies galaxies with
particles that lie above a sharp threshold density in the final conditions, 
and
one that associates galaxies statistically with high peaks 
of the initial
density field.  For the $b=1.5$ CDM model we also try a somewhat more
sophisticated approach, applying a galaxy identification algorithm
that has been ``calibrated'' from cosmological simulations with hydrodynamics.

This last scheme is based on the simulation of Cen \& Ostriker 
(1993, hereafter CO; see also Cen \& Ostriker 1992), and we refer to it
as `C/O biasing.'  CO simulate the standard CDM model ($\Omega=1$, $\Gamma=0.5$)
with $\sigm=0.77$ and a simulation volume 80 $h^{-1}$ Mpc on a side.
Their simulations include dissipation and star formation in a baryonic
component, and they apply a percolation algorithm to group ``star'' particles
into ``galaxies'' at $z=0$.  In their analysis, they fit the relation 
between the smoothed galaxy number density $\ng$ and the smoothed mass density
$\rhom$ to the functional form
$$
\log(\ng/\ngavg) = A + B\log(\rhom/\rhomavg) + 
	      C\left[\log(\rhom/\rhomavg)\right]^2.
\eqno(1)
$$
CO list values of $A$, $B$, and $C$ for Gaussian smoothing filters of 
various radii.  We want to select particles from an $N$-body simulation,
and for this purpose it is easier to use cubic cells 
rather than
Gaussian filters, since every particle is a member of a distinct cell.
R.\ Cen has kindly computed the parameters of equation (1) for us
using 2 $h^{-1}$ Mpc cubic cells.  To select biased particles, we compute
the final mass density field on a $150^3$ grid and apply equation (1) to 
compute the galaxy density in each cell, using a mean 
galaxy density $\ngavg$ equal to the density of VGH's densest observational
sample (for which $\dbar \equiv \ngavg^{-1/3} = 4.5$ $h^{-1}$ Mpc).
We then select particles from cell $\vec{x}$ with probability 
$p=\ng(\vec{x})/\np(\vec{x})$, where $\np(\vec{x})$ is the particle
number density in the cell.
We use cloud-in-cell weighting to
compute the density field and to assign selection probabilities to particles.
For 2 $h^{-1}$ Mpc cells, Cen finds $B=1.9$ and $C=-0.20$; the value of
$A$ is determined by the requirement that the mean density of selected
particles equal the desired mean galaxy density $\rhogavg$.
As a check, we have computed 
galaxy and mass density fields from our biased and unbiased particle
distributions, with Gaussian smoothing filters of radius
3 and 5 $h^{-1}$ Mpc.  The relation between smoothed galaxy density and
smoothed mass density is then close to equation (1), as evaluated 
with the values of $B$ and
$C$ that CO report for these smoothing radii.  This agreement confirms
that our particle selection scheme provides a reasonable representation
of CO's biasing results.

Of the three biasing schemes that we use, the C/O scheme has the clearest
physical motivation.  However, one should bear in mind that the CO simulations
have a resolution of only $\sim 400$ $h^{-1}$ kpc, which may not be adequate 
for identifying galaxies reliably.  The simulations of Katz, Hernquist \&
Weinberg (1992) and Evrard, Summers \& Davis (1993), which model the baryon
component using smoothed-particle hydrodynamics, have much higher spatial
resolution in galaxy-forming regions, but the simulation volumes are not
large enough to allow a meaningful definition of a biasing relation like
equation (1).  For now, we take CO's result as a plausible illustration of
the sort of biasing relation that can be derived from cosmological simulations
with gas dynamics and galaxy formation.  Since the C/O relation is
derived specifically for standard CDM with $b\approx 1.5$, we apply it only to
this particular model.  Changing the normalization, the shape of the power
spectrum, or the value of $\Omega$ would alter the collapse times 
of galaxy-scale perturbations, so it is not at all obvious
how the properties of biasing would respond to these changes.

Our `density biasing' scheme is similar to the C/O scheme discussed above, but
instead of the non-linear function (1), we simply adopt a sharp threshold ---
galaxies form with equal probability where the smoothed mass density exceeds
some value, and they do not form at all where the mass density is below this
value.  Specifically, we create a final density field from 
the $z=0$ distribution of mass points, and smooth it with
a Gaussian filter of radius $R_{\rm s}$.  We set $R_{\rm s}=2.25$
$h^{-1}$ Mpc for consistency with the peaks biasing procedure described
below. We then compute the smoothed density at the final location of
each particle, and select some fraction $f_{\rm b}$ of particles that have
the highest local densities. As this fraction diminishes, the
corresponding subset of particles becomes more and more biased towards
the densest regions of the simulation. We randomly sample this biased
subset to the desired inter-galaxy separation $\overline d = 4.5$ $h^{-1}$ Mpc, 
and we identify the
resultant population of points as the `galaxies' of interest. The
value of $f_{\rm b}$ is chosen so $\sigma_{8,{\rm gal}}(t_0)$
for this biased and sampled subset matches that of the corresponding
unbiased simulation.  This sharp threshold in final density is the biasing
scheme used by EEGS and by WC.

The `peaks biasing' scheme, which has been widely used in previous
studies of CDM models, is simple in conception but complicated to
implement in practice.  One might naturally expect that peaks of
the primordial density field would become the first sites of non-linear 
gravitational collapse (though Katz, Quinn \& Gelb [1993] find a rather poor
correspondence between galaxy-scale peaks of the initial density field
and collapsed dark halos in $N$-body simulations).
The assumption behind peaks biasing is that bright galaxies form only at
galaxy-scale peaks of the initial density field that lie above some
global threshold $\nu _{\rm t}\equiv \delta _{\rm mass}/\sigma
_{\rm mass}>1$.  Lower height peaks are assumed to
form underluminous galaxies or structures not recognized as galaxies at
all.  A number of authors have discussed physical mechanisms that
might lead to such a thresholding effect, among them
Rees (1985), Silk (1985), Dekel \& Silk (1986), White {\it et al.\/} (1987),
Kaiser (1988), and Cole \& Kaiser (1989).

Kaiser (1984) showed that the high peaks of a Gaussian density field cluster
more strongly than the underlying mass distribution, and he suggested
that this phenomenon might explain the high amplitude of the observed
cluster-cluster correlation function.  BBKS applied the same idea to
galaxy formation, and they computed many of the statistical and clustering
properties of peaks of Gaussian fields.
The strength of biasing between the high peaks and the underlying density 
field depends on
the importance of long wavelength perturbations in the initial
conditions. Long wavelength perturbations modulate the background
density, and they therefore alter the local effective threshold height,
\ie the relative ease with which a peak relative to the {\it local\/}
background can rise above a {\it global\/} threshold. In a Gaussian
random field, the number density of rare peaks is a strong
function of peak height, so small changes in the effective
threshold height can lead to large changes in the local number density
of peaks above a global threshold.  If the primordial fluctuations have
appreciable large-scale power (as in CDM), then the high peaks occur
preferentially in regions of high background density. 
Bright galaxies that form around these high peaks are `born' clustered,
so they provide biased tracers of the underlying mass distribution.

When applied to a given mass distribution, the peaks biasing scheme
has two free parameters. The first is the smoothing scale
for defining galaxy peaks, $R_{\rm g}$ (in our case, $R_{\rm g}$ 
is the radius of a Gaussian window function). This radius defines
the characteristic physical size of the peaks, so it
should correspond approximately to the mass scale of a typical galaxy.
Since the precise relation between the smoothing scale and the
collapsed mass is uncertain, we have a fair amount of freedom in our
choice of $R_{\rm g}$. The second free parameter is the peak threshold
height $\nu_{\rm t}$ -- this measures the `difficulty' of
galaxy formation. The values of $R_{\rm g}$ and $\nu _{\rm t}$ can be
fixed by requiring that the biased model with which they are
associated fit two observational numbers, usually taken to be
the galaxy number density (or the associated characteristic separation
$\overline d\equiv n^{-1/3}$, in our case $4.5$ $h^{-1}$ Mpc) 
and the rms amplitude of galaxy count fluctuations on some scale (in our case,
$\sigma _{8,{\rm gal}}\simeq 1$). 

Our initial conditions do not resolve galaxy scales, so we implement
peaks biasing via the peak-background split approximation, which allows
one to identify particles with galaxy-scale peaks in a statistical 
manner, given the density field smoothed over a larger scale
(BBKS; see also Park 1991).  The approximation relies on the fact that
$\delta _{\rm g}$ -- the full initial density field smoothed
on the galaxy scale $R_{\rm g}$ -- has the same power spectrum and hence
the same statistical properties as the field
$\delta _{\rm b} + \delta_{\rm p}$, where $\delta_{\rm b}$ is a field
with the same underlying power spectrum as $\delta_{\rm g}$ but smoothed
on a larger, background scale $R_{\rm b}$, and $\delta_{\rm p}$ is a `peaks'
field whose power spectrum is that of
$\delta _{\rm g}$ minus that of $\delta _{\rm b}$. 
One can therefore estimate the local density of peaks in $\delta_{\rm g}$
above a global threshold $\nu_{\rm t}$ as the number of peaks in
$\delta_{\rm p}$ above a local effective threshold, whose value is modulated
by the background field $\delta_{\rm b}$.  The use of the peak-background
split approximation introduces a third parameter into our biasing
prescription, the background smoothing length $R_{\rm b}$.
Given our choice of $R_{\rm g}$ (see
below), we adopt a Gaussian window radius $R_{\rm b}=2.25 h^{-1}$ Mpc. 
This value ensures that $3<R_{\rm b}/R_{\rm g}<5$, as required for best 
results in the peak-background split approximation (BBKS, Park 1991). 
With this condition satisfied, our results should not be sensitive to
the exact value of $R_{\rm b}$. 

In all our implementations of peaks biasing, we use equation (6.35) of
BBKS to calculate the expected local number density 
$n _{\rm pk}(\delta_{\rm b})$ of
galaxy-scale peaks above $\nu_{\rm t}$, as a function of the background
field height $\delta_{\rm b}$. 
We identify a mass particle that lies in a given cell $\vec{x}$ of the
initial conditions as a `galaxy' 
with a probability equal to 
$A \cdot V \cdot n_{\rm pk}[\delta_{\rm b}(\vec{x})]$,
where $V$ is the cell volume and $A$ is a constant of proportionality
discussed below.  At a mechanical level, this operation is quite similar
to that in density biasing or C/O biasing, except that we select particles
based on the initial rather than the final density field, and we use the
rather complicated relation between galaxy and mass density implied by
the peaks formulae instead of a sharp threshold function or the 
C/O relation.  In a strict implementation of peaks biasing,
the proportionality constant $A$ is equal to one, and in this case
the biasing operation has a clear physical interpretation:
the selected particles correspond, statistically, to galaxy-scale peaks
of the initial density field.  To compare our models with VGH's brighter,
less dense
samples (those with $\dbar>4.5$ $\hmpc$), we randomly sample our biased particle
distributions to the desired number density.  By so doing, we make the
implicit assumption that galaxy biasing is independent of luminosity, at least
for galaxies brighter than those of VGH's densest sample (absolute 
magnitudes $M<-18.5$).  We could instead have identified brighter galaxies
with higher, rarer peaks, which are more strongly clustered.  The first
assumption makes physical sense if random factors independent of peak 
height determine a galaxy's final luminosity.

At a fixed value of $R_{\rm g}$, the strength of biasing is quite sensitive
to the threshold $\nu_{\rm t}$.  One might therefore think that the peaks
biasing prescription could yield any bias factor of interest with an
appropriate choice of parameters, in particular that for any value
$1/3 \simlt \sigma_{8,{\rm mass}} \simlt 1$ one could find values of
the peak parameters $\nu_{\rm t}$ and $R_{\rm g}$ that would yield
$\sigma_{8,{\rm gal}}=1$.  However, there is a second observational constraint
to be matched, in our case the characteristic separation 
$\dbar = 4.5$ $\hmpc$.  We find that with the galaxy density or separation 
imposed as a constraint, the strict peaks biasing prescription 
(with the proportionality constant $A=1$) can yield only
a rather narrow range of bias factors for a given mass fluctuation spectrum.

This limitation arises because $R_{\rm g}$ and $\nu_{\rm t}$ both
affect $\overline d$ and $\sigma _{8,{\rm gal}}$ in the same sense.
If the smoothing length $R_{\rm g}$ is fixed, then
increasing $\nu_{\rm t}$ increases $\overline d$ (since higher peaks are
rarer), and it increases $\sigma _{8,{\rm gal}}$ (since
higher peaks are more strongly clustered). If the threshold height
$\nu_{\rm t}$ is fixed, then increasing $R_{\rm g}$ increases $\overline
d$ (since it reduces the choppiness of the density field and thus
increases the space between peaks), and it
increases $\sigma _{8,{\rm gal}}$ for a reason that is
readily appreciated in terms of the peak-background split. 
Raising $R_{\rm g}$ reduces the amplitude of fluctuations associated with 
the `peaks' field $\delta_{\rm p}$, so the large-scale waves
have more influence in raising local peaks above the global threshold,
making the global peaks more strongly biased towards regions of high
background density (see BBKS equation 6.35).
With the strict peaks biasing
scheme, raising $\nu_{\rm t}$ to get a higher bias factor
also raises $\overline d$. To keep $\overline
d$ fixed (as we require for our comparison to the VGH data), $R_{\rm
g}$ must be correspondingly lowered, but this also lowers $b$.  
These two effects on $b$ nearly cancel in our models, so over the plausible
range of peaks parameters the bias factor in a given model is
nearly constant.

For our $\Gamma=0.5$ and $\Gamma=0.25$ models, the strict peaks biasing
prescription works well for $\sigma_{8,{\rm mass}}=1/1.5$ ($b=1.5$).
The peak parameters $R_{\rm g}=0.57$ $\hmpc$ 
and either $\nu_{\rm t}=1.8$ (for $\Gamma=0.5$) or $\nu_{\rm t}=1.6$ 
(for $\Gamma=0.25$) yield $\sigma_{8,{\rm gal}} \approx 1$ and 
$\overline d \approx 4.5$ $h^{-1}$ Mpc, as desired.
However, when $\sigma_{8,{\rm mass}}=1/2$ or $1/3$, we cannot find peak
parameters that produce a strong enough bias, \ie if we require 
$\dbar=4.5$ $\hmpc$, then no choice of peak parameters gives 
$\sigma_{8,{\rm gal}}=1$.
We therefore adopt a relaxed version of the peaks prescription that
can produce higher bias factors.
To avoid the cancellation effect described above, we fix
$R_{\rm g}$ at $0.57$ $\hmpc$, and we vary $\nu_{\rm t}$ to obtain the
desired bias factor (\ie to obtain $\sigma_{8,{\rm gal}}=1$).
We then vary the proportionality constant $A$ in order to achieve the
desired characteristic separation, $\dbar=4.5$ $\hmpc.$
For the $b=2$ and $b=3$ models we require $A>1$, making the abundance of 
selected galaxy particles larger than the abundance of peaks.
Although our biasing formula is based on the peaks approach, in these cases
there is no direct correspondence between the selected particles and peaks.
This is somewhat unsatisfying, but since the peaks model (and any other
biasing model) is at best an approximation of physical processes whose details
are not well understood, it seems sensible to loosen the model a bit
when necessary.  For $b=2$ the values of $\dbar$ that we would obtain
with $A=1$ are $\sim 6-7$ $\hmpc$, so for the two brightest VGH samples,
which have $\dbar=7.4$ and $10.9$ $\hmpc$ respectively, we can still identify
the simulations' galaxy particles with a subset of the high peaks.

\sectionbegin {3. Results} 

\subsectionbegin {3.1 Visual Appearance of the Models} 

Figure 2 displays representative Cartesian slices from the final-time
particle distributions of our models. 
Each panel displays a 15 $\hmpc$ thick slice through the simulation
cube, with the galaxy population sampled to the characteristic
separation $\overline d = 5.6$ $h^{-1}$ Mpc of VGH's $M_{\rm lim}=-19$
samples.  Axis scales are marked in $\hmpc$.
Figure 2a displays 
slices from unbiased and density-biased power-law models
(simulation cubes 192 $\hmpc$ on a side).
Its three rows 
show 
$n=0$, $-1$, and $-2$ models from top to bottom,
and its four columns show 
$b=1$, 1.5, 2, and 3 models from left to right.
Figures 2b and 2c display galaxies from the \gam\ models
(simulation cubes 300 $h^{-1}$ Mpc on a side).
The columns have the same $b$ associations as Figure 2a,
but Figures 2b and 2c display
peaks-biased and
density-biased models, respectively.
The three rows of these figures are associated with
($\Gamma$, $\Omega_0$, $\lambda_0$) = 
(0.5, 1, 0), (0.25, 1, 0), and (0.25, 0.3, 0.7) models, from top to bottom. 
Throughout
Figure 2 the
largest rounded voids are much smaller than the panels,
indicating that our simulation cubes are
indeed large enough to allow a fair measure of voids
for the range of models that we consider.

\topinsert
\capskip
\centerline{
\epsfxsize=6.5truein
\epsfbox[36 306 594 738]{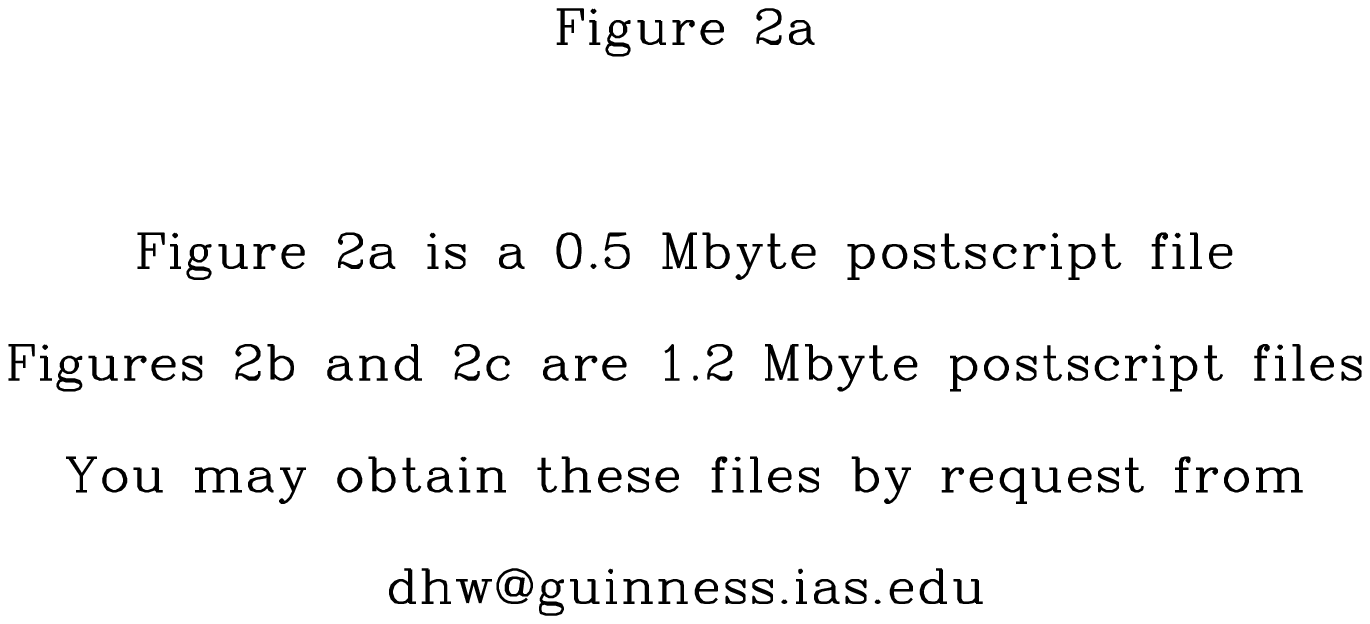}
}
\medskip
{\eightpoint
\noindent {\bf Figure 2} --- 
Slices through the simulated galaxy distributions at $z=0$.
In every panel the rms galaxy fluctuation in spheres of radius
8 $h^{-1}$ Mpc is unity, the characteristic inter-galaxy separation is
$\overline d$ = 5.6 $h^{-1}$ Mpc, and the thickness of the Cartesian 
slice is 15
$h^{-1}$ Mpc. For all three plots, the four columns from left to right show
models with bias factors $b$ = 1, 1.5, 2, and 3.
(a) Unbiased and density-biased power-law models, with 
$n$ = 0, $-1$, and $-2$ in the rows from top to bottom.
(b) Unbiased and peaks-biased models, with 
($\Gamma$, $\Omega_0$, $\lambda_0$) = (0.5, 1, 0), (0.25, 1, 0), and 
(0.25, 0.3, 0.7)
from top to bottom.
(c) Same as (b), but with peaks-biased slices replaced 
by corresponding density-biased slices.
}
\medskip
\textskip
\endinsert

\topinsert
\capskip
\centerline{
\epsfxsize=6.5truein
\epsfbox[36 306 594 738]{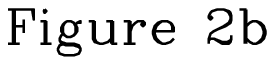}
}
\medskip
{\eightpoint
\noindent {\bf Figure 2} --- {\it continued}
}
\medskip
\textskip
\endinsert

\topinsert
\capskip
\centerline{
\epsfxsize=6.5truein
\epsfbox[36 306 594 738]{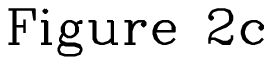}
}
\medskip
{\eightpoint
\noindent {\bf Figure 2} --- {\it continued}
}
\medskip
\textskip
\endinsert

The initial conditions of all the
models in Figure 2a 
were generated using the same random phases,
as were all the models in Figures 2b and 2c.
Thus within Figure 2a and throughout Figures 2b and 2c,
recognizable structures tend to form at similar locations in different
slices. Each model nonetheless produces distinctive structure, and one
can notice several trends. 
Comparing panels within a given row of Figure 2, 
as one moves from left to right
the low density regions become emptier, and the
high density regions become more diffuse.
Thus biasing has two basic
effects on spatial structure: it turns low density regions into
completely empty voids, and it reduces the non-linearity of high density
regions. The first effect arises because biasing suppresses
galaxy formation in low density regions.
The second effect arises because 
the mass fluctuations in our biased models
are smaller, and they are
therefore less efficient at inducing gravitational collapse. 
Recall that all of the models pictured in Figure 2 have the same rms fluctuation
in galaxy counts at 8 $\hmpc$.

Comparing corresponding $b\geq 1.5$ panels from Figures 2b and 2c, it is clear
that the empty voids in density-biased models are larger than their 
peaks-biased counterparts. 
[Correspondingly, we will find that
the density-biased VPFs in our simulations are always higher
than their peaks-biased counterparts.]
This is to be expected, since there is no more
efficient way of creating voids than by `density biasing', \ie
by completely eliminating all particles from low density
regions of the final conditions. 
Peaks biasing, on the other hand, permits occasional galaxy formation
in low density regions.
The rounded,
`bubble-like' voids in many of the biased models 
are reminiscent of features in wedge diagrams from the CfA2 redshift survey 
(de Lapparent \etal 1986; 1991) or the Perseus-Pisces redshift survey
(Haynes \& Giovanelli 1986; 1988).  
However, one should be cautious in drawing conclusions from visual
comparisons, because properties of real-space, Cartesian slices 
like those in Figure 2
can be quite different from those of the redshift-space, magnitude-limited, 
declination wedges that are often used to display redshift survey results.

Our spectra are normalized to have the same amplitude
at $\sim 8$ $h^{-1}$ Mpc.  In the power-law models of Figure 2a, the
models with steeper spectra (lower $n$) have more power
on larger scales and less power on smaller scales. Both of these
properties manifest themselves clearly in the slices.
Models with steeper spectra
develop coherent features that can be traced over a larger
fraction of the simulation cube, and models with shallower spectra develop
final structure that is clumpier on small scales. 
The \gam\ models  of Figures 2b and 2c
exhibit less variation in spatial structure 
than the power-law models
because the differences in the initial power spectra are themselves much
smaller.  However,
comparing the $\Omega_0=1$ models in Figures 2b and 2c, $\Gamma=0.5$ in
the top rows and $\Gamma=0.25$ in the middle rows, one sees 
that void sizes increase slightly among biased models as 
$\Gamma$ is lowered. It is also possible to detect smoother,
more coherent filaments in the models with more large-scale power
(lower $\Gamma$).  Comparing the middle and bottom rows in the same
figures shows the effect of lowering $\Omega_0$ while keeping the
initial power spectrum fixed.
Lowering $\Omega _0$ 
reduces peculiar velocities on all scales because there is less mass 
available to attract condensations away from pure Hubble flow.
However, $\Omega_0$ has little effect on final spatial structure if
fluctuations are normalized to the final epoch --- at the level of
the Zel'dovich approximation or the adhesion approximation it has
no effect at all (see Weinberg \& Gunn 1990).
Some systematic differences between high-$\Omega_0$ and low-$\Omega_0$
models appear in the fully non-linear regime (see WC), but at the
limited resolution of these slice plots, the differences are essentially
undetectable.
In terms of visual appearance, the effects of biasing easily outweigh
the effects of $\Omega_0$ or of the initial power spectrum.

\subsectionbegin {3.2 Measuring the VPF}

Our principal measure of the size and frequency of voids is the void
probability function, the probability $P_0(R)$ that a
randomly placed sphere of radius $R$ contains no galaxies. We also
employ a related statistic, the probability $P_{80}(R)$ that the average
density in a randomly placed sphere is more than $80\%$ below the global
mean density.  We also call $P_{80}(R)$ the underdense region 
probability function, or UPF.  
White (1979) discussed the relation between count probability statistics,
including the VPF, and correlation function statistics.
Following White's original suggestion,
many later studies of the VPF adopted the `scaled' variables
$\chi(\bar N \bar \xi) \equiv -{\rm ln}[P_0(R)]/{\bar N}$, where
$\bar N$ and $\bar \xi$ are respectively the mean number of galaxies and
the mean value of the two-point correlation function in spheres of 
radius $R$
(\eg Fry 1986, Maurogordato \& Lachi\`eze-Rey 1987,
Fry \etal 1989; Lachi\`eze-Rey, da Costa \& Maurogordato 1992).
This choice of variables allows one to test the so-called hierarchical
hypothesis, the assumption that all
higher-order correlation functions can be expressed as sums of pairwise
products of two-point correlation functions. 
However, the introduction of $\bar \xi$ into the scaled radius variable
obscures the relation between the scaled VPF, $\chi(\bar N \bar \xi)$, and the 
actual sizes of voids.
Since we are interested in the frequency of large voids rather than the
hierarchical hypothesis, we have decided to stick with the simpler 
representation $P_0(R)$.

VGH present VPFs for volume-limited subsets of the CfA redshift survey
with the absolute magnitude limits $M_{\rm lim}=-18.5$, $-19.0$,
$-19.5$, and $-20.0$ (for $h=1$).  
The characteristic separations corresponding to these four
limits,
derived from the galaxy luminosity function, are
$\overline d = 4.5$, 5.6, 7.4, and 10.9 $h^{-1}$ Mpc, respectively.
The procedures described in \S 2 create simulated galaxy samples
with a final-time separation of $\overline d =4.5$ $h^{-1}$ Mpc.
To compare our predictions with all the
data of VGH, we subsequently sample this population to separations of
$\overline d = 5.6$, 7.4, and 10.9 $h^{-1}$ Mpc.
As mentioned in \S 2, because we randomly sample to compare to brighter
subsets of the VGH data, our models implicitly assume that biasing (if any)
is independent of galaxy luminosity for $M<-18.5$.

For our power-law models, we present VPF results only in real space,
ignoring possible distortions from peculiar velocities.  Measuring the VPF
in real space is simple because our simulation volume is triply periodic.
We choose 2000 random points throughout the simulation volume,
and about each point we count the number of galaxies in concentric
spheres of radius 1, 2, 3, $\ldots$ $R_{\rm max}$ $h^{-1}$ Mpc, 
including particles across a periodic boundary if necessary.
We adopt $R_{\rm max}=35$ $\hmpc$, a radius by which $P_{80}(R)$ drops
to zero in nearly all of our models.
We set $P_0(R)$ equal to the number of empty spheres of radius $R$ divided
by the total number (2000) of spheres of radius $R$ placed in the simulation
volume.  Similarly, we set $P_{80}(R)$ equal to the number of spheres that
are at least 80\% underdense, \ie that contain 
$N \leq 0.2(4\pi R^3/3\bar d^3)$
galaxies, divided by the total number of spheres of radius $R$.
We have four independent simulations of each power-law model, and the VPFs
that we plot are the average over these four runs.

For our \gam\ models we evaluate the VPF and UPF in real space and in redshift
space, \ie using the 
spherical coordinates ($r$, $\theta$, $\phi$)
and ($r+v_r/H_0$, $\theta$, $\phi$), where $v_r$ is the radial peculiar
velocity with respect to the observer, and $H_0$ is the Hubble constant.	
Adopting a $v_r$-dependent coordinate breaks the assumption of
periodic boundary conditions, so our procedure for evaluating the VPF
is more complicated than before.  We want to measure the VPF in spherical
rather than cubical samples because their redshift-space distortions are
more likely to mimic those of a real observational sample --- cubes have
corners where the sample is unusually deep in the radial direction.
We also want our samples to fill the simulation cube, so that we take
full statistical advantage of the simulation volume.  In each 300 $\hmpc$
simulation cube, therefore, we select 8 `observers' located at the 
vertices of a smaller cube of side $L/2=150$ $\hmpc$.  For each observer
in turn, we shift particles (using the periodic boundaries) so that the
observer lies at the center of the cube.  We then move galaxies to redshift
space and select a spherical sample containing all galaxies within 
a redshift distance $R_{\rm sample}$ = 130 $h^{-1}$ Mpc of the observer.
We choose 500 random points within a sphere of radius 
$(R_{\rm sample}-R_{\rm max})$, and about each point we count galaxies in 
concentric spheres of radii 1, 2, 3, $\ldots$ $R_{\rm max}$ $h^{-1}$ Mpc. 
We define $P_0(R)$ and $P_{80}(R)$ as before, the number of empty or
underdense spheres divided by the total number (500).
We measure the real-space VPF and UPF in the same fashion, except that
we omit the shift from real space to redshift space.
All of a simulation's cubical volume appears in at least one of the cube's
8 samples, and most of it appears in more than one.
There are two independent simulations of each \gam\ model, and the results
that appear in our 
figures are the average over all 16 (= $2\cdot 8)$ relevant samples.

Figure 3 compares the real-space (solid line) and redshift-space 
(dotted line) VPFs of
our unbiased, $(\Gamma, \Omega_0, \lambda_0)$ = (0.25, 1, 0) model.
[Following WC, we
use $\overline d=5.6$ $h^{-1}$ Mpc here and in any other figures that
display the VPF for only one value of $\overline d$.]
At a given radius, the void probability tends to be slightly higher
in redshift space than in real space, probably
because coherent
outflows from low density regions move galaxies outward from the
positions that they occupy in physical space, thereby increasing the
sizes of voids along the line-of-sight (cf. figure 6 of 
Reg\"os \& Geller 1991).
Although
redshift-space VPFs 
in our simulations are almost always higher than 
their
real-space counterparts, there are 
a few rare instances where they are lower -- especially at large radii.
Lower redshift-space VPFs can arise because
velocity dispersions in groups and clusters 
scatter galaxies outward into regions that are empty in physical space.
The differences between real-space and redshift-space VPFs are generally
smaller than those in Figure 3 for models 
with smaller $\Omega_0$,
larger $\Gamma$,
larger $b$,
or smaller $\overline d$.
The trends with $\Omega_0$, 
$\Gamma$,
and 
$b$
are comprehensible, in that decreasing the mass density
or the scale or amplitude of mass fluctuations
leads directly to 
a decrease in large-scale peculiar 
velocities.
The trend with $\overline d$ probably
occurs
because the coherent outflows associated with voids
are less evident at smaller separations.

\topinsert
\capskip
\centerline{
\epsfysize=3.0truein
\epsfbox[72 410 540 738]{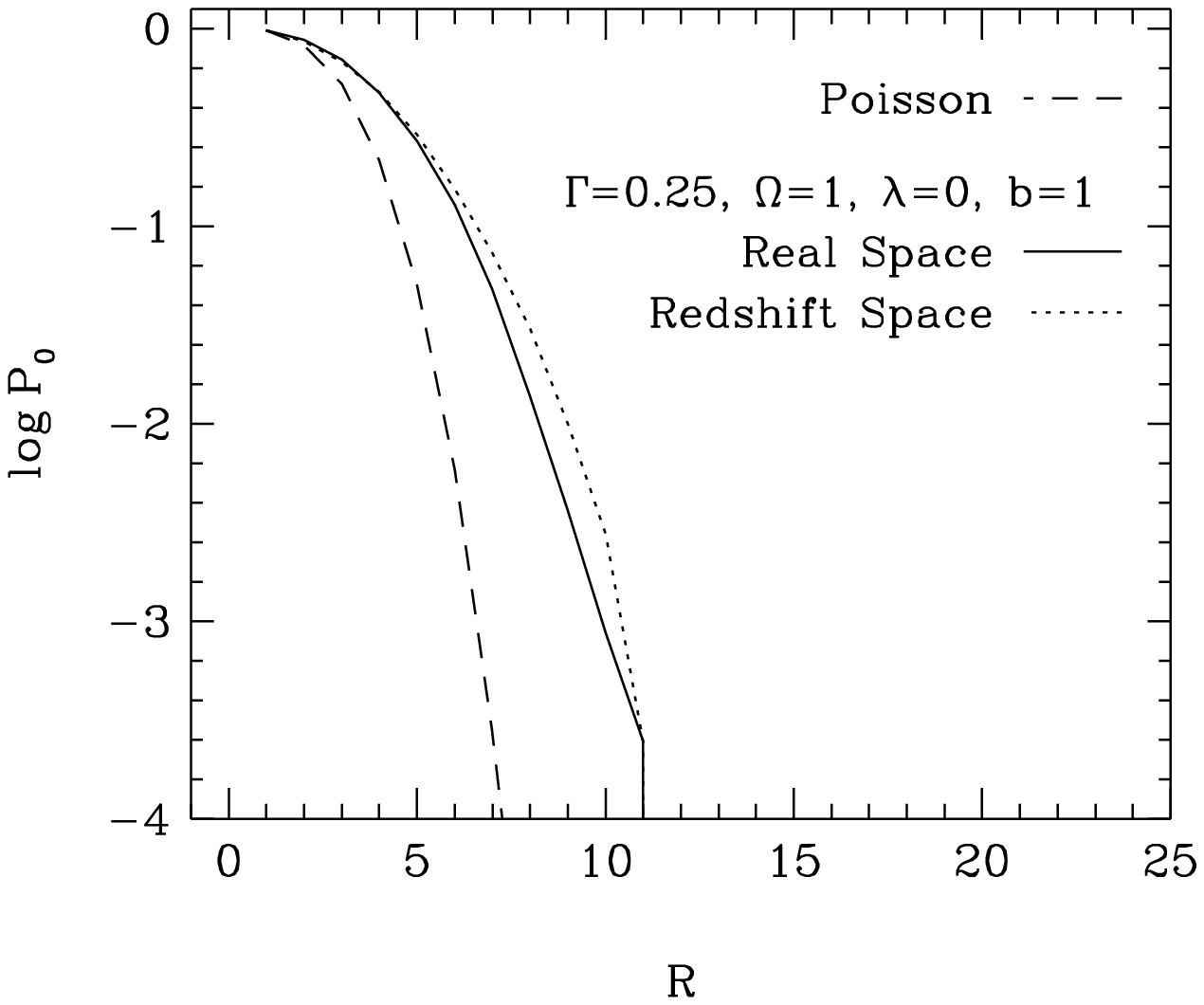}
}
{\eightpoint
\noindent {\bf Figure 3} --- 
The void probability function (VPF) in real space and redshift space.
$P_0(R)$ is the
probability that a randomly placed sphere of radius $R$ (in $h^{-1}$
Mpc) is empty of galaxies. 
Solid and dotted curves mark the real-space and redshift-space VPFs
of our unbiased, ($\Gamma$, $\Omega_0$,
$\lambda_0$) = (0.25, 1, 0) model, with $\bar d=5.6$ $\hmpc$.
VPFs tend to be slightly higher
in redshift space than in real space, probably because
coherent outflows from
low density regions shift galaxies outward from the positions that
they occupy in physical space,
stretching voids along the line-of-sight.
The dashed curve shows the mean VPF for a Poisson distribution of
particles with the same value of $\bar d$, 
$P_0(R)={\rm exp}(-4\pi R^3/3{\bar d}^3)$.
VPFs of the $N$-body models lie far above the Poisson VPF at large radii,
as expected.
}
\medskip
\textskip
\endinsert

The dashed curve in Figure 3 shows the VPF expected for a Poisson
distribution of points with characteristic separation 
$\bar d$,
$P_0(R)={\hbox{exp}}(-4\pi R^3 / 3 \bar d^3)$.
The model VPFs lie far above the Poisson VPF at large radii, as expected.
Nonetheless, in Figure 3 and in the VPFs of all of our other models,
$P_0$ drops to $10^{-4}$ at radii much smaller than $L/2$ = 150 $h^{-1}$ Mpc,
confirming
Figure 2's indication that even our largest voids are much
smaller than the simulation cube.

\subsectionbegin {3.3 Dependence of the VPF on Model Parameters} 

\subsubsectionbegin {3.3.1 Unbiased Models} 

\topinsert
\capskip
\centerline{
\epsfysize=3.0truein
\epsfbox[72 410 540 738]{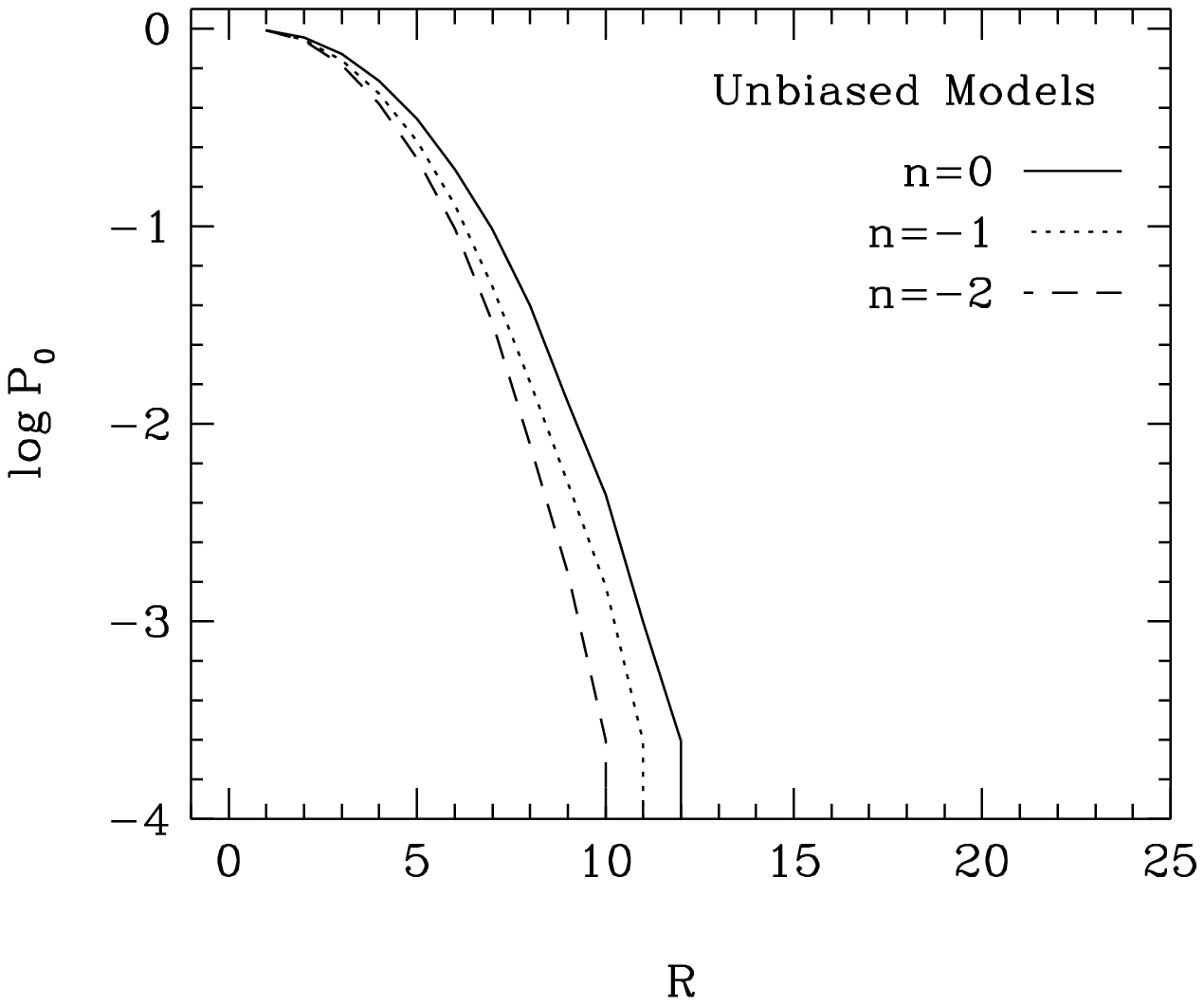}
}
{\eightpoint
\capskip
\noindent {\bf Figure 4} --- 
Dependence of the VPF on the initial power spectrum, for unbiased models.
Solid, dotted, and dashed lines show the real-space VPFs of unbiased
power-law models with $\bar d=5.6$ $\hmpc$ and $n=0$, $-1$, and $-2$, 
respectively.  Increasing $n$ leads to a modest increase in the VPF,
but the trend is rather weak.
}
\endinsert
\textskip

Figure 4 illustrates the dependence of the VPF
on the shape of the initial
power spectrum
for unbiased models.
Solid, dotted, and dashed lines represent the real-space VPFs 
for our unbiased power-law models
with $n=0$, $-1$, and $-2$, respectively. 
The three curves are quite similar, showing that 
the VPF in unbiased models depends only weakly on the shape of the 
initial power spectrum.
This insensitivity probably indicates that large empty regions 
grow from fluctuations near the 8 $\hmpc$ normalization scale, where all
three models have the same rms amplitude.
The VPFs in this figure nonetheless exhibit a systematic trend --
as $n$ is lowered, the VPF decreases slightly. The same trend is
visible in the corresponding slice plots (column one of Figure 2a).
WC also noted this effect among Gaussian models,
and they cited two reasons for it. First,
introducing small-scale power (increasing $n$)
creates stronger negative
fluctuations on small scales, and these lead directly to higher VPFs on
those scales.
Second, introducing small-scale power
clumps galaxies on small scales, reducing the effective
number of independent tracers; since the effective inter-particle
density is thus lowered, the VPF goes up.
Models with lower $n$ have stronger fluctuations on large scales, but
on these scales the fluctuation amplitude may be too low to 
clear out empty regions.

Figure 5 illustrates
the dependence of the VPF (or lack thereof) on 
$\Omega_0$ and 
$\lambda_0$, for unbiased models.
Five curves are plotted:
solid, dotted, short-dashed, long-dashed, and 
dot-dashed lines for the real-space VPFs
of unbiased, $\Gamma=0.25$ models with the parameter combinations
($\Omega_0$, $\lambda_0$) = (1, 0), (0.3, 0), (0.1, 0), (0.3, 0.7), and
(0.1, 0.9), respectively.  Four of the curves lie so close to each other
that they are indistinguishable on this plot.  Only the 
($\Omega_0$, $\lambda_0$) = (0.1, 0) VPF stands slightly apart, and even
in this case the differences in the tail of the VPF correspond
to just one or two voids out of the 8000 randomly placed spheres.
Figure 5 demonstrates that the real-space VPF is extremely insensitive
to $\Omega_0$ and $\lambda_0$, if the power spectrum is held fixed.
This insensitivity is not terribly surprising, since the
final spatial structure is completely independent of 
$\Omega_0$ and $\lambda_0$
at the
level of the adhesion approximation, and one expects dynamical non-linearities
to distinguish these models mainly in collapsed, virialized 
regions.
In redshift space we find a mild dependence of the VPF on $\Omega_0$ because
the magnitude of peculiar velocity distortions depends on $\Omega_0$.
However, we have already seen that the difference between real-space and
redshift-space VPFs is small (Figure 3), and the differences between 
redshift-space VPFs with the same power spectrum but different values of
$\Omega_0$ are smaller still.  At fixed $\Omega_0$, even the redshift-space VPF
is insensitive to $\lambda_0$ because peculiar velocities depend almost
entirely on the density parameter rather than the cosmological constant.

\topinsert
\capskip
\centerline{
\epsfysize=3.0truein
\epsfbox[72 410 540 738]{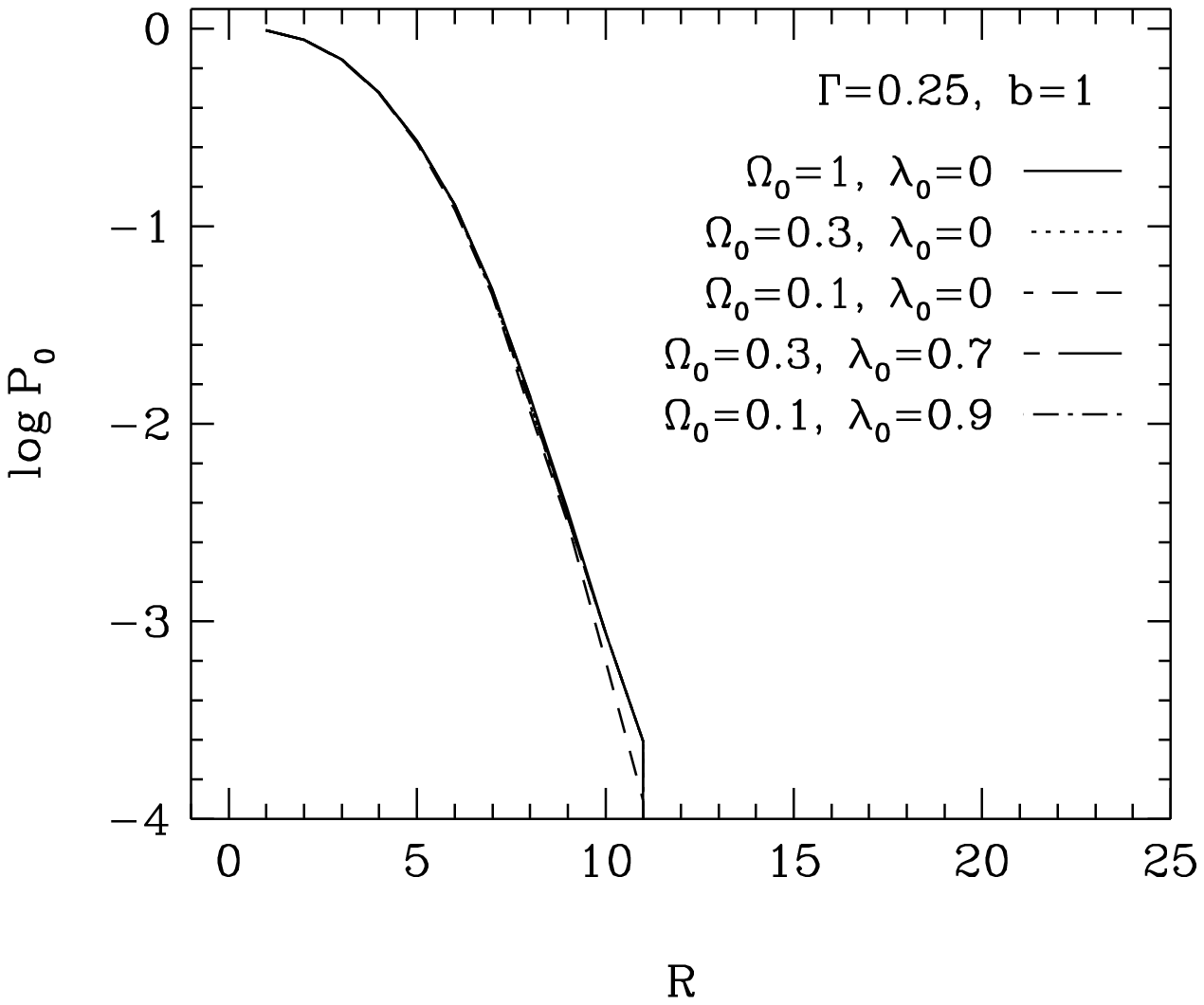}
}
{\eightpoint
\capskip
\noindent {\bf Figure 5} --- 
Dependence of the VPF
on
$\Omega_0$ (the cosmic density parameter) and
$\lambda_0$ (the cosmological constant).
Different line types show real-space,
$\overline d$ = 5.6 $h^{-1}$ Mpc
VPFs of five unbiased, $\Gamma$ = 0.25
models with 
five different combinations of 
$\Omega_0$ and
$\lambda_0$, as labeled.
Four of these VPFs are so similar that the curves are indistinguishable
on this plot, and the fifth ($\Omega_0=0.1$, $\lambda_0=0$) is only
marginally different.  Within the parameter range shown here,
$\Omega_0$ and $\lambda_0$
have essentially no impact on the real-space VPF. 
The value of $\Omega_0$ has a small effect on the redshift-space VPF, since it
determines the amplitude of peculiar velocities.
}
\endinsert
\textskip

\vskip 1 truecm
\subsubsectionbegin {3.3.2 Biased Models} 

Figure 6 illustrates the dependence of the VPF on the prescription adopted
for biased galaxy formation, at a fixed bias factor $b=1.5$ and a
characteristic separation $\bar d=5.6$ $\hmpc$.
Solid, dashed, and dotted lines display real-space VPFs for standard CDM
models ($\Gamma=0.5$, $\Omega_0=1$), biased to $b=1.5$ with
peaks, density, and C/O biasing, respectively.
Density biasing produces much higher VPFs than peaks biasing, as one
would expect given the visual appearances of Figures 2b and 2c.
This difference between the two biasing schemes holds for all of our 
models; it is even more pronounced for $\Gamma=0.25$ 
(more large-scale power).
Peaks biasing and C/O biasing, on the other hand, produce nearly identical VPFs.
Although the difference between these schemes is somewhat larger for other
values of $\bar d$, it is always very small.

\topinsert
\capskip
\centerline{
\epsfysize=3.0truein
\epsfbox[72 410 540 738]{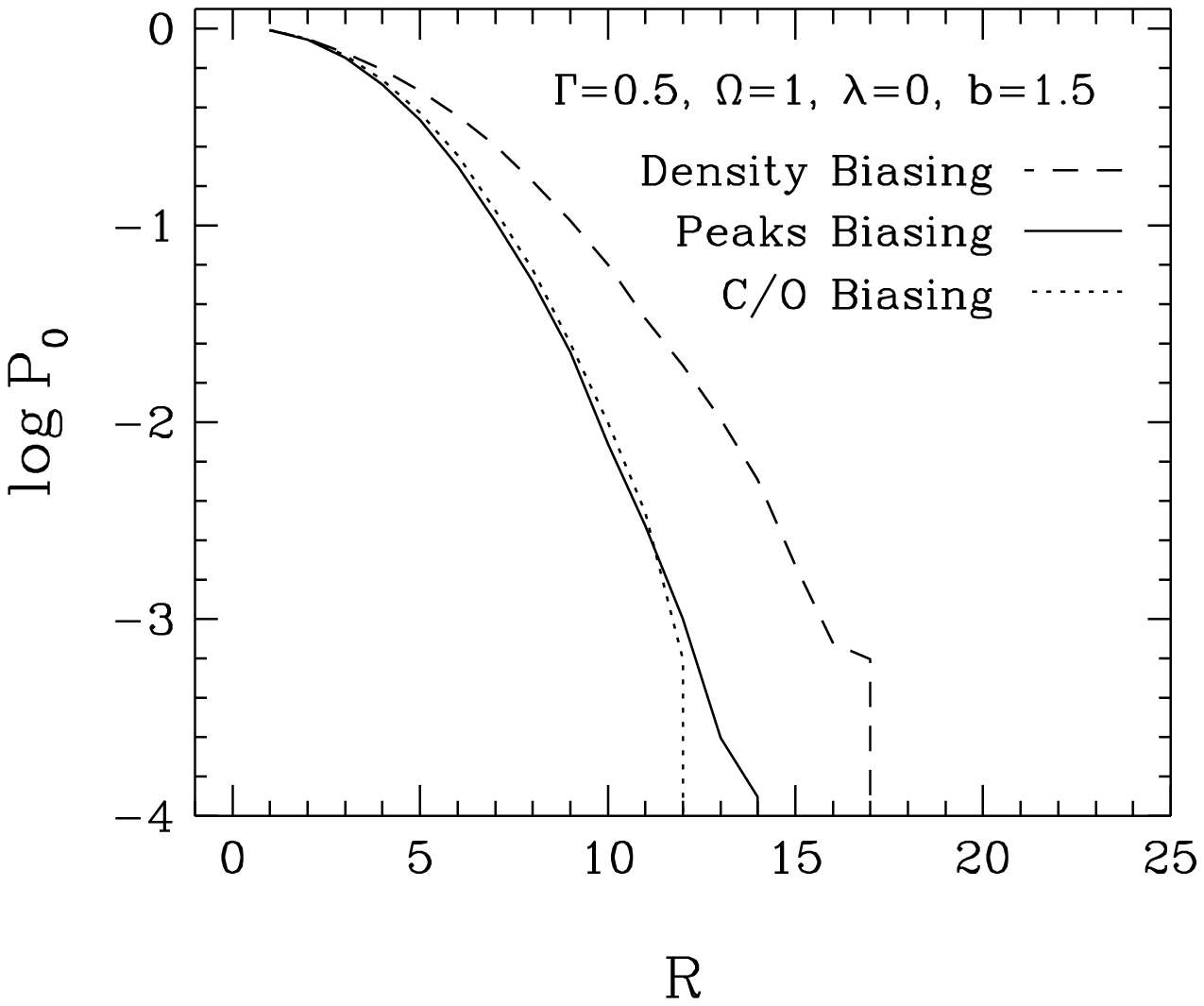}
}
{\eightpoint
\noindent {\bf Figure 6} --- 
Dependence of the VPF on the prescription for biased galaxy formation,
at a fixed value of the bias factor $b$.
Solid, dashed, and dotted lines show real-space, $\bar d=5.6$ $\hmpc$ VPFs
for $b=1.5$, `standard CDM' models with peaks biasing, density biasing,
and C/O biasing, respectively.
Density biasing creates larger voids and a higher VPF than peaks biasing;
the same trend holds for other models, other bias factors, and other
values of $\bar d$.  Peaks biasing and C/O biasing yield similar VPFs.
}
\medskip
\line{
\epsfysize=2.9truein
\epsfbox[108 410 468 738]{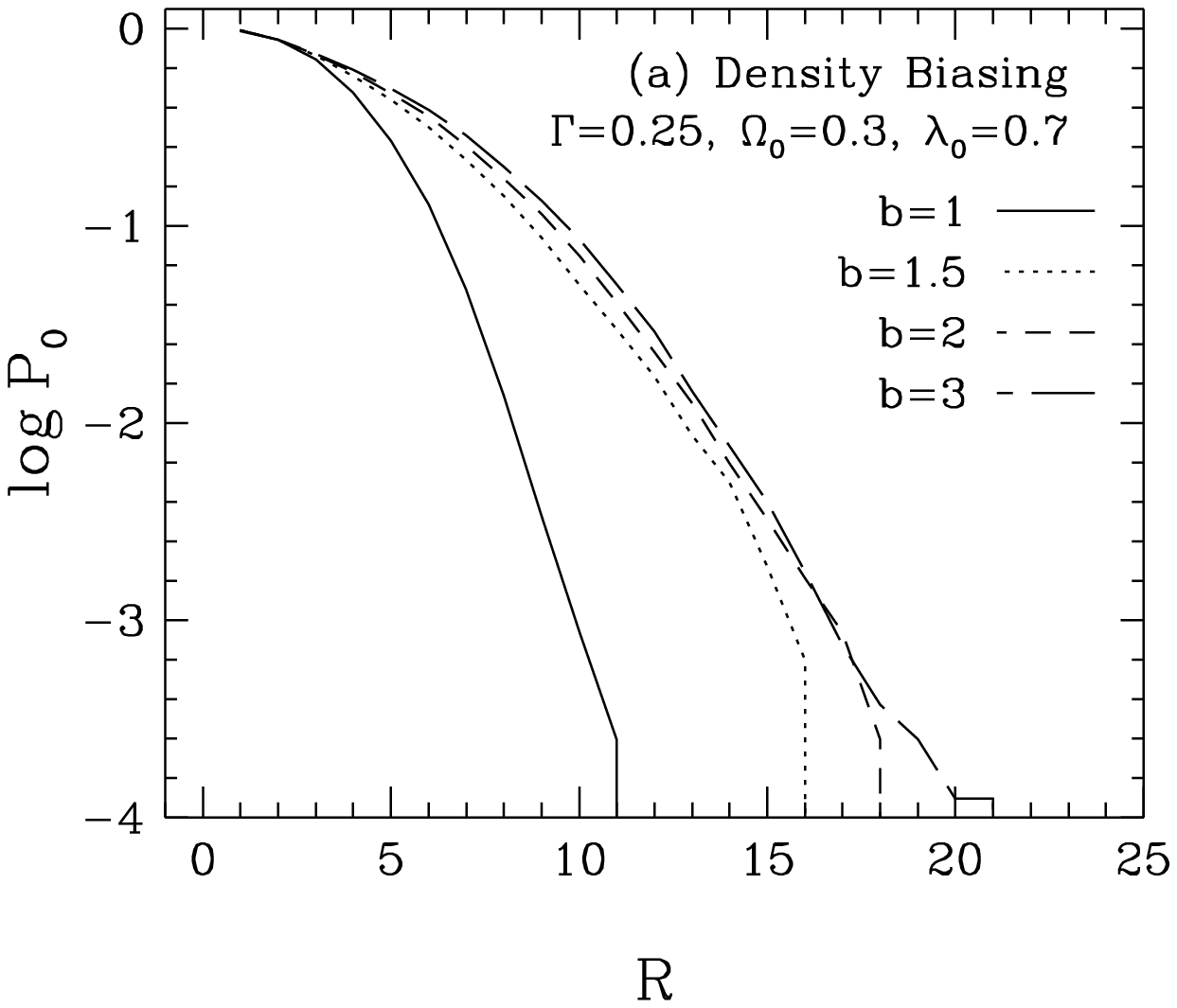}
\hfil
\epsfysize=2.9truein
\epsfbox[108 410 468 738]{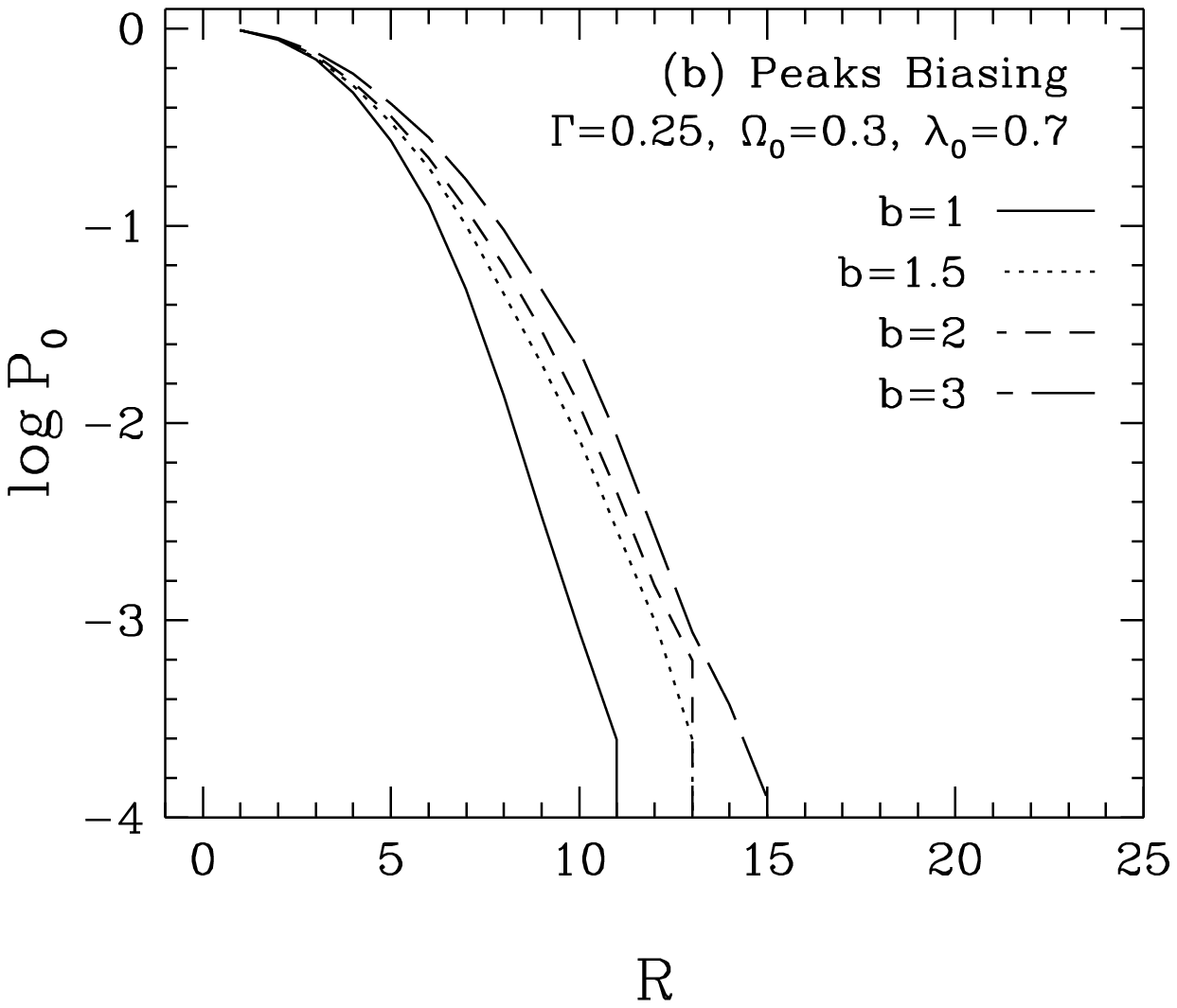}
}
{\eightpoint
\capskip
\noindent {\bf Figure 7} --- Dependence of 
the VPF on the bias factor $b$.
(a) Real-space,
$\overline d$ = 5.6 $h^{-1}$ Mpc
VPFs for 
unbiased and density-biased models with
$\Gamma$ = 0.25,
$\Omega_0$ = 0.3, and $\lambda_0$ = 0.7.
The VPF jumps sharply between 
$b=1$ and $b=1.5$, but it increases only slowly over
the entire remaining range, from $b=1.5$ to $b=3$. This 
behaviour contrasts with that in EEGS's figure 4b, to which this plot
can be compared.
(b) Same as (a), but
with density-biased VPFs replaced 
by corresponding peaks-biased VPFs.
The latter are more like the corresponding unbiased VPF, though again
there is a larger jump between $b=1$ and $b=1.5$ than between higher
bias factors.
}
\endinsert
\textskip

Figure 7 illustrates the dependence of the VPF on the bias factor $b$,
for density biasing (Figure 7a) and peaks biasing (Figure 7b).
Figure 7a is designed for comparison with figure 4b of EEGS, which
also shows real-space VPFs for 
spatially flat, low-$\Omega_0$, CDM models with
different bias factors.  Except for our choice of smoothing filter and
our dilution to a fixed galaxy density, our density biasing scheme is
the same as EEGS's biasing scheme.
In Figure 7a, solid, dotted, short-dashed, and long-dashed lines represent
real-space VPFs of our 
$(\Gamma$, $\Omega_0$, $\lambda_0)$ = (0.25, 0.3, 0.7) models, with biasing
factors of $b=1$, 1.5, 2, and 3, respectively.
In both EEGS's figure 4b
and our Figure 7a,
increasing the bias factor raises the VPF.
However, the form of this trend differs
dramatically in the two cases.
In EEGS's figure, four successive increases in the bias factor
produce a steady and uniform march of the VPF to
appreciably higher values. 
EEGS's median void radii display a similar steady trend
(see their table 2 and figure 6).
It is exactly this systematic dependence which suggested that the VPF
might be a sensitive indicator of biasing, and perhaps a statistical tool
with which to determine the bias factor.
In our Figure 7a, however, the VPF
increases only weakly over the entire range $1.5\leq b \leq 3$,
despite a large increase between $b=1$ and $b=1.5$.
This result is consistent with the visual appearance of successive
columns in Figure 2c.

There are many differences between our simulations and those of EEGS,
but the difference in normalization procedure is probably the most
important for explaining the different dependence on the bias factor.
For their CDM models, EEGS take a fixed underlying mass distribution
and choose successively more strongly biased subsets of the particles.
Each increase in the bias factor yields a `galaxy' population that is
more strongly clustered in an rms sense, and it is not surprising that
void sizes increase each time.  However, at most one of these biased
subsets can match the known rms fluctuation of galaxy counts as measured
from the two-point correlation function.  We adopt the observed rms
fluctuation at 8 $\hmpc$ as a constraint on our `galaxy' populations, so
we accompany each increase in the bias factor with a corresponding decrease
in the amplitude of the underlying mass fluctuations.  
The reduction in mass fluctuations counteracts the stronger biasing; in an
rms sense the two effects cancel, by construction.
Void sizes are sensitive to the efficiency of galaxy formation in low
density regions, so some dependence of the VPF on $b$ remains, but this
dependence is strong only between $b=1$ and $b=1.5$.

Figure 7b illustrates the dependence of the VPF on the bias factor for 
the peaks biasing prescription.
It is obtained 
by replacing the
density-biased VPFs of Figure 7a 
with their peaks-biased counterparts. 
Once again
the VPF is more sensitive to the difference between $b=1$ (unbiased) and
$b=1.5$ than to the differences between $b=1.5$, 2, and 3. 
However, the jump from $b=1$ to $b=1.5$ is much smaller than it is for
density biasing, and the growth of the VPF
with $b$ is somewhat more steady.

\topinsert
\capskip
\centerline{
\epsfysize=3.0truein
\epsfbox[72 410 540 738]{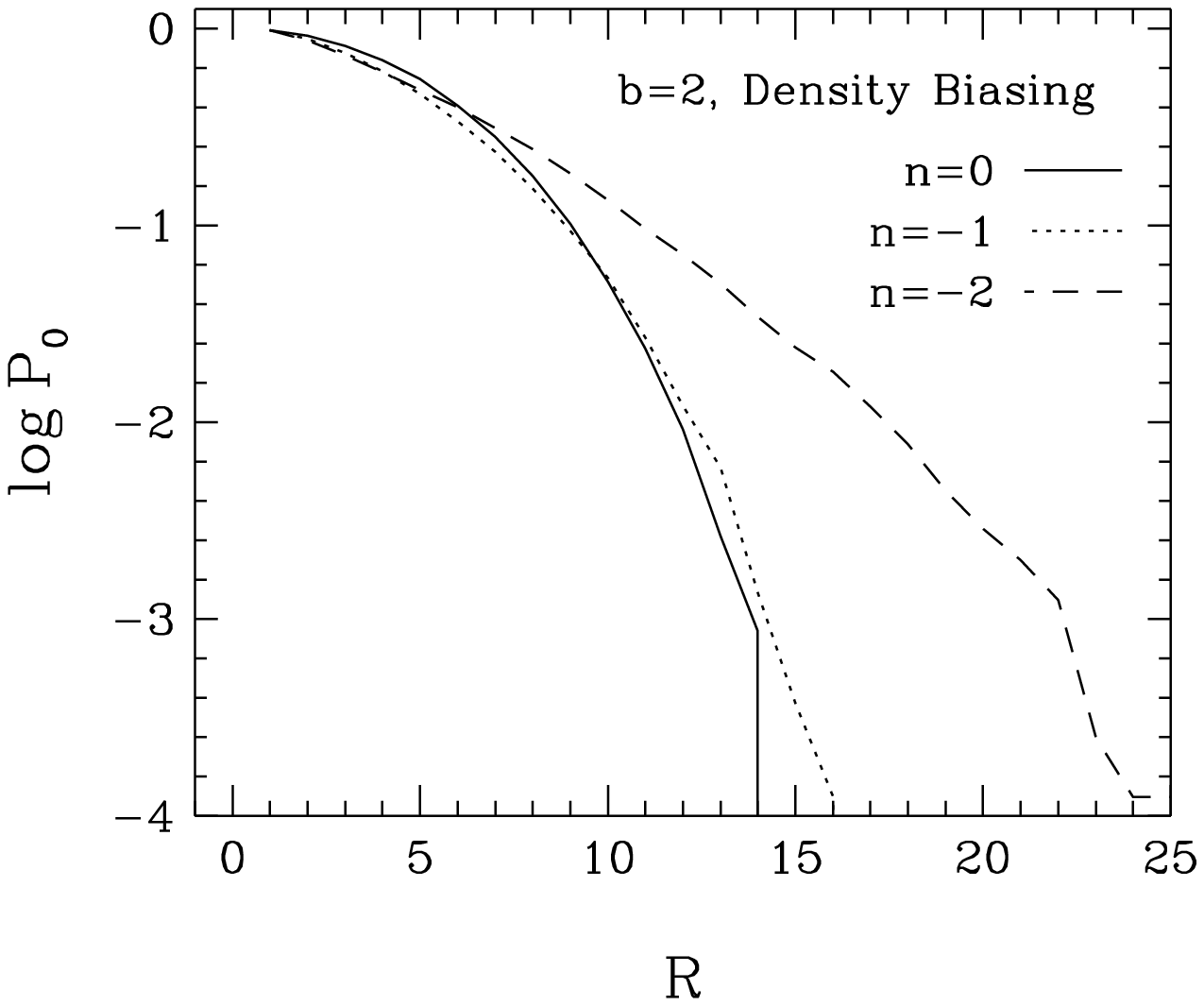}
}
{\eightpoint
\noindent {\bf Figure 8} --- 
Dependence of the VPF on the initial power spectrum, for biased models.
Solid, dotted, and dashed lines show the real-space VPFs of density-biased
power-law models with $\bar d=5.6$ $\hmpc$ and $n=0$, $-1$, and $-2$, 
respectively.  The first two VPFs are similar, but the biased $n=-2$ VPF
is much higher, as biasing turns the $n=-2$ model's large underdense
regions into empty voids.  The trend for biased models is quite
different from that for unbiased models (cf. Figure 4).
}
\medskip
\textskip
\endinsert

Figure 8 illustrates the dependence of the VPF 
on the shape of the initial
power spectrum 
for biased models
(analogous to Figure 4 for unbiased models).
Solid, dotted, and dashed lines display the real-space
VPFs of our $b=2$,
density-biased, power-law models with
$n=0$, $-1$, and $-2$, respectively. 
The $n=0$ and $n=-1$ models have similar VPFs, but the $n=-2$ VPF
is much higher.  This behaviour contrasts with that in Figure 4, 
where all three models have similar VPFs and the $n=-2$ VPF is the lowest.
WC found similar results for their biased models.
While the large-scale fluctuations in the $n=-2$ models are too weak to
clear out empty voids by gravity alone, they do create large regions
of low mass density, and biasing turns these regions into large, empty voids.
We find the same trend --- larger voids in biased models with more
large-scale power --- when comparing our $\Gamma=0.5$ and $\Gamma=0.25$ models,
using either density biasing or peaks biasing.  If one knew that
galaxy formation were biased and one knew the specific nature of the bias,
one could conceivably 
derive constraints on the primordial power spectrum from the VPF.
However, given the uncertainties in the appropriate choice of biasing
scheme, the VPF is probably not a useful diagnostic for $P(k)$.
Voids tell us more about the relation between galaxies and mass
than about the power spectrum of the mass distribution itself.

\subsectionbegin {3.4 Basic Comparison to the Data} 

The Vogeley, Geller \& Huchra (VGH) void probability data come from 
three magnitude-limited redshift surveys, `CfA1 North', `CfA2 North',
and `CfA2 South'.  The `CfA1 North' survey consists of the northern
part of the original CfA survey (Huchra \etal 1983), 
which is complete to a limiting apparent magnitude of 14.5, covers the
range $\delta > 0^\circ$, $b^{II}>40^\circ$, and contains 1833 galaxies.
The `CfA2 North' survey comes from the extension of the CfA survey to
the limiting magnitude 15.5 (see Geller \& Huchra 1989).  It covers the
range $8^h \leq \alpha \leq 17^h$, $26.5^\circ \leq \delta \leq 44.5^\circ$,
and it contains 2556 galaxies.  The `CfA2 South' survey, also from the
CfA extension, contains 2414 galaxies in the range $20^h \leq \alpha \leq 4^h$,
$6^\circ \leq \delta \leq 36^\circ$.  
From each of these three surveys, VGH construct four volume-limited samples,
the absolute magnitude of the faintest galaxies in these samples being
$M_{\rm lim} = -18.5$, $-19.0$, $-19.5$, and $-20$ (for $h=1$).  
The corresponding characteristic inter-galaxy
separations, computed from the galaxy luminosity function,
are $\overline d = 4.5$, 5.6, 7.4, and 10.9 $h^{-1}$ Mpc, respectively
(see table 1 of VGH).
These 12 volume-limited samples contain between 182
and 627 galaxies each, and they
range in depth from 40 to 126 $h^{-1}$ Mpc. 

VGH computed VPFs for each of these 12 observational samples, and 
M.\ Vogeley has kindly provided these data to us in the form of computer files.
Figure 9 compares the observational data to the predictions of our 
\gam\ models.
Open circles, asterisks, and open triangles display the VPFs measured
from `CfA2 North', `CfA2 South', and
`CfA1 North', respectively.
The four rows 
display VPFs for $\overline d = 4.5$, 5.6, 7.4, and 10.9
$h^{-1}$ Mpc samples, from top to bottom.
The same 12 observational VPFs appear in each column
of panels in Figure 9.
The trend toward larger
VPFs from top to bottom within a given column reflects the simple fact
that increasing the characteristic inter-galaxy separation (decreasing
the galaxy density) increases the sizes of empty voids. 
We have not attached error bars to the points of Figure 9; we will consider
the effects of finite volume errors in the next section.
For now we note simply that the three, largely independent observational
samples give fairly consistent but not identical results.

\pageinsert
\capskip
\centerline{
\epsfysize=6.8truein
\epsfbox[0 0 612 774]{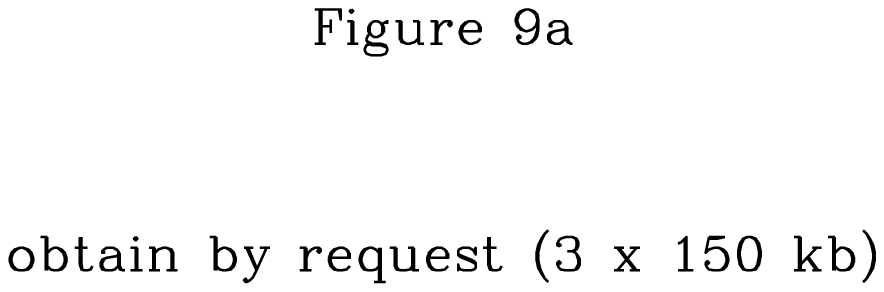}
}
{\eightpoint
\capskip
\noindent {\bf Figure 9} --- 
Comparison between observed VPFs and the VPFs of our best-motivated 
theoretical models.
Within each panel:
({\it i\/}) the circles, 
asterisks, and triangles are the observed, redshift-space
VPFs calculated by Vogeley, Geller, and Huchra (1991; VGH) from 
volume-limited samples of the `CfA2 North',
`CfA2 South', and `CfA1 North' galaxy redshift surveys, respectively;
({\it ii\/}) the solid, dotted, 
short-dashed, and long-dashed lines are redshift-space
VPFs calculated from our theoretical models
with $b=1$, 1.5, 2, and 3, respectively. Within each of the three plots:
({\it i\/}) the four rows from top to 
bottom show VPFs for characteristic inter-galaxy 
separations $\overline d$ = 4.5, 5.6, 7.4, and 10.9 $h^{-1}$ Mpc;
({\it ii\/}) 
the $b\geq 1.5$ lines in the left and right columns are associated with
peaks biasing and density biasing, respectively.
The 
dot-dashed lines in the right column of (a) are the redshift-space
VPFs from our
$b=1.5$, C/O-biased model.  
Figures (a), (b), and (c) show model VPFs for
($\Gamma$, $\Omega_0$, $\lambda_0$) = 
(0.5, 1, 0), (0.25, 1, 0), and (0.25, 0.3, 0.7), respectively.
These VPFs exhibit many of the model-dependent trends discussed earlier.
Unbiased, peaks-biased, and C/O-biased models generally reproduce the
VGH data fairly well, while density-biased models tend to create an 
excess of large voids.

}
\vfill
\endinsert
\textskip

\topinsert
\capskip
\centerline{
\epsfysize=6.8truein
\epsfbox[0 0 612 774]{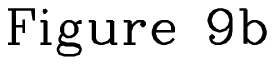}
}
{\eightpoint
\capskip
\noindent {\bf Figure 9} --- {\it continued}
}
\endinsert
\textskip

\topinsert
\capskip
\centerline{
\epsfysize=6.8truein
\epsfbox[0 0 612 774]{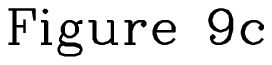}
}
{\eightpoint
\capskip
\noindent {\bf Figure 9} --- {\it continued}
}
\endinsert
\textskip

Figures 9a, 9b, and 9c compare the VGH results to the
redshift-space VPFs
of our
$(\Gamma$, $\Omega_0$, $\lambda_0$) = 
(0.5, 1, 0),
(0.25, 1, 0), and
(0.25, 0.3, 0.7) models, respectively.
We do not include a comparison to the power-law models because they
have less observational and theoretical motivation; WC show some comparisons
between these models and the VGH data.
In each panel of Figure 9, the
solid, dotted, short-dashed, and
long-dashed lines are associated with
$b=1$, 1.5, 2, and 3, respectively.
Left hand columns show results for peaks biasing, right hand columns for
density biasing (the solid lines show unbiased models in each case).
The dot-dashed lines in the right hand column of Figure 9a are
associated with the C/O biasing scheme, for $b=1.5$.
Figure 9 exhibits many of the model trends seen in previous figures. For 
example,
it confirms that 
density biasing produces significantly larger VPFs than
peaks biasing,
and that peaks biasing and C/O biasing produce nearly identical VPFs
(cf. Figure 6).
Figure 9 also confirms 
that the VPFs of unbiased models
depend only weakly on
$\Omega_0$, 
$\lambda_0$, and the initial power spectrum
(cf. Figures 4 and 5), and
that increasing large-scale power can 
substantially increase
density-biased
VPFs
(cf. Figure 8) --
as $\Gamma$ is lowered, Figure 9's density-biased
VPFs
increase appreciably.
This effect
is also present, though less pronounced, 
among peaks-biased models.
Finally,
within a given panel of Figure 9
the difference between $b=1$ and $b\geq 1.5$ VPFs is often much larger than
the difference among $b\geq 1.5$ VPFs (cf. Figure 7).
This effect 
is 
especially strong among density-biased
models and models
with more large-scale power, because they 
are least like their unbiased counterparts.
The
effect is weaker for larger values of $\overline d$ because increasing
$\overline d$ raises unbiased VPFs relative to their biased
counterparts. 
The impact of increasing $\bar d$ is evidently greater when
a smaller fraction of the total volume
is already empty (\ie in unbiased models). 

Turning to a comparion with the data, it is clear that
the density-biased models of Figure 9 generally tend to overproduce 
voids. 
This tendency
weakens (and occasionally reverses)
as $\overline d$ increases (\ie as the 
observed
VPFs go up), or as
large-scale power or $\Omega_0$ decrease (so that model VPFs go down).
However,
all
of our density-biased models match the data less well than 
their peaks-biased or C/O-biased counterparts. 
The slice plots of density-biased models (Figure 2) share the `bubbly'
appearance of the CfA2 slices, but the model voids are considerably
larger than those in the CfA2 survey.

Most of the unbiased, peaks-biased, and C/O-biased models of Figure 9 
match the VGH data fairly well.
Unbiased models typically fit best at small $\bar d$, while peaks-biased
or C/O-biased models fit better at larger $\bar d$, with $b \approx 1.5-2$
models perhaps the most successful overall.
Among
unbiased and peaks-biased models,
the data do not seem to strongly prefer any one of our three
$(\Gamma$, $\Omega_0$, $\lambda_0$) = (0.5, 1, 0), (0.25, 1, 0), 
(0.25, 0.3, 0.7) parameter
combinations. 
The high-$\Gamma$ models fit better at $\bar d=5.6$ $\hmpc$, but they
underproduce large voids at $\bar d=10.9$ $\hmpc$, where the low-$\Gamma$
models are more successful.
All of our biased models
overestimate the VPF at
$\overline d = 5.6$ $h^{-1}$ Mpc, while
all of our unbiased models underestimate the VPF at
$\overline d =10.9$ $h^{-1}$ Mpc, so no one model offers 
an ideal fit to the data. 
However, we have not yet considered the effects of finite volume errors,
so we cannot draw quantitative conclusions.

\subsectionbegin {3.5 Error Estimates for Specific Models} 

Although the CfA2 survey is one of the deepest and most ambitious redshift
surveys to date, the total volume that it probes does not exceed the volume
of the largest low density regions by an enormous factor.  Estimates of
the VPF are therefore subject to significant finite volume errors ---
a distant observer mapping a volume of the universe of the same size
might find a 
different VPF because of statistical fluctuations in the local structure.
These finite volume effects dominate over other sources of error
(\eg magnitude and redshift errors) in the observed VPF.
To decide whether a particular theoretical model is consistent with
the VGH data, we must ask whether observers mapping equal volumes
of the model universe could reproduce the observed VPF a reasonable
fraction of the time.  

The best way to carry out such a comparison between simulations and data
is to draw from the numerical models simulated data sets with the same 
geometry and selection effects as the observational survey, then analyze
the simulated and real data in an identical fashion.  We do not know precisely
what procedures VGH used for defining their samples and placing their
test spheres, so here we adopt the simpler procedure of analyzing spherical
samples with the same volumes as the VGH samples.
This method should still yield reasonable estimates of the finite
volume fluctuations for each sample.
Within each of the two simulations of a given 
model, we
place 50 `observers' (\ie origins) at random positions.
Around each observer in turn, we move the particles into redshift space
and measure the VPF in eight spherical samples, with volumes chosen to
match those of VGH's eight
volume-limited CfA2 samples.
The radii
$R_{\rm sample}$ of these spheres range
from 22.9 $h^{-1}$ Mpc (giving the volume of the 
$\overline d$ = 4.5 $h^{-1}$ Mpc,
CfA2 North sample)
to 54.7 $h^{-1}$ Mpc (giving the volume of the 
$\overline d$ = 10.9 $h^{-1}$ Mpc,
CfA2 South sample). 
In each sample sphere we randomly select a fraction 
(4.5 $\hmpc/\bar d_{\rm s})^3$ of
the particles, where $\bar d_{\rm s}$ is the characteristic separation of the
associated CfA2 sample, so that on average the mean galaxy density in 
our samples matches that of observed galaxies at the associated
absolute magnitude limit.
We measure the redshift-space VPF by counting galaxies in spheres of
radius 1, 2, ..., $R_{\rm max}$ $\hmpc$, placed at 8000 random positions out
to a distance ($R_{\rm samp} - R_{\rm max}$) from the observer.
We adopt $R_{\rm max}=12$, 14, 16, and 22 $\hmpc$ for the four successive
values of $\bar d$.  These values are large enough to go beyond the largest 
voids found by VGH at each $\bar d$.  The larger $R_{\rm max}$, the smaller
the volume we have in which to place test spheres, so we always attempt to keep
$R_{\rm max}$ significantly smaller than $R_{\rm sample}$.

Our procedure produces 100 (= 50 observers per realization
$\times$ 
$2$ realizations)
measures of the VPF
for each of the eight CfA2 sample volumes, and we sort these 100 measures from
the lowest to the highest VPF at each radius $R$. 
The 5th, 25th, 75th, and 95th 
positions in this ranking represent our best estimates of the 
5\%, 25\%, 75\%, and 95\% VPFs associated with a given model.
At a given radius, one out of 20 observers measures a VPF value lower than
the 5\% VPF, and one out of 20 measures a VPF higher than the 95\% VPF.
Half of the observers measure a VPF between the 25\% and 75\% levels.
If the observational data fall in this inter-quartile range, then the
model matches the data well.  If the data fall in the 5\%$-$95\% range,
then the model and data agree at the ``$2\sigma$'' level.
Successive points on the VPF are not independent, since a large void at
one radius must contain voids at all smaller radii. 
VPFs at different $\bar d$ are also not independent because a void that
is empty at one $\bar d$ must be empty at all larger $\bar d$.
These interdependences make it difficult to combine multiple VPF
measurements for multiple samples into overall model likelihoods.
We will not attempt to solve this statistical problem in this paper;
we leave the task of devising useful likelihood tests to a more
detailed comparison study that applies matched procedures to 
simulations and observations.

We cannot attach ``error bars'' directly to the observational data in
a model-independent way because we do not know the underlying statistical
distribution from which 
the real data are drawn.
The expected finite-volume errors are different for each theoretical model ---
models with larger voids also have larger sample-to-sample fluctuations
in the VPF --- so we must compare the data individually to each model in turn.
Figures 10a, 10b, and 10c show such a comparison for three cases
that span the range of our \gam\ model VPFs:
the unbiased, ($\Gamma$, $\Omega_0$, $\lambda_0$) = (0.25, 0.3, 0.7) model,
the $b=1.5$, C/O-biased, ($\Gamma$, $\Omega_0$, $\lambda_0$) = (0.5, 1, 0) 
model, and the
$b=3$, density-biased, ($\Gamma$, $\Omega_0$, $\lambda_0$) = (0.25, 1, 0) 
model.  The first of these 
is a representative unbiased model,
the second is arguably the most 
physically motivated 
of our biased 
models, and the third creates the largest voids of all our \gam\ models.
Left and right hand columns of Figure 10 are
associated with the sample volumes of 
`CfA2 North' and `CfA2 South',
respectively, and
the four rows correspond to 
$\overline d = 4.5$, 5.6, 7.4, and 10.9 $h^{-1}$ Mpc,
from top to bottom.
Circles (left hand columns) and asterisks (right hand columns) 
show the VGH data.
The pair of solid lines in each panel shows the 5\% and 95\% VPFs of
the corresponding model; the region between them can be regarded as a
``2$\sigma$ error corridor'' for the model.
The pair of dotted lines shows the 25\%
and 75\% VPFs.
Both pairs of lines diverge as $P_0$
descends because of the logarithmic scale on the $P_0$-axis.

\pageinsert
\capskip
\centerline{
\epsfysize=6.8truein
\epsfbox[0 0 612 774]{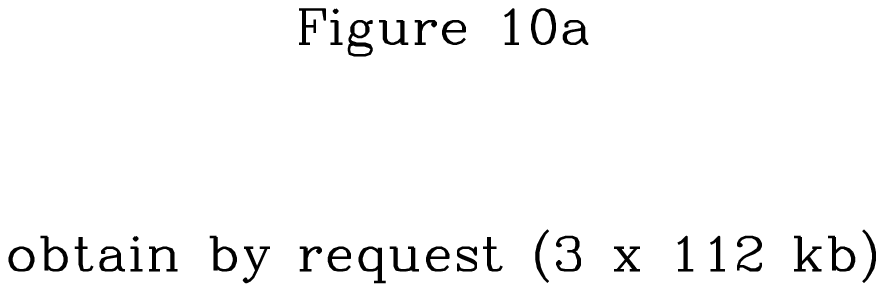}
}
{\eightpoint
\capskip
\noindent {\bf Figure 10} --- 
Detailed comparison between three of our models and the VGH data,
including the effects of finite volume fluctuations.
Circles (left panels) and asterisks (right panels) show the VGH data
from `CfA2 North' and `CfA2 South,' respectively, with 
$\bar d = 4.5$, 5.6, 7.4, and 10.9 $\hmpc$ from top to bottom.
For each of these eight samples, we measure the VPF from 100 simulated
data sets drawn from the theoretical model, each with the same volume as
the corresponding observational sample.
Solid lines in each panel mark the 
5th-lowest and 5th-highest VPFs of the 100 model data sets;
the region between them can be regarded as a
``2$\sigma$ error corridor'' for the model.
Dotted lines mark the 25th-lowest and 25th-highest VPFs, and thus
represent the model's inter-quartile range.
(a) The unbiased, ($\Gamma$, $\Omega_0$, $\lambda_0$) = (0.25, 0.3, 0.7) model.
The data are consistent with the model at the ``$2\sigma$'' level, except 
for the large-$R$ tail in the northern, $\bar d=4.5$ $\hmpc$ sample.
(b) The $b=1.5$, C/O-biased, ($\Gamma$, $\Omega_0$, $\lambda_0$) = (0.5, 1, 0) 
model.  This model provides an excellent fit to the data overall, though again
there is a small area of discrepancy in the tail of the northern, 
$\bar d=4.5$ $\hmpc$ sample.
(c) The $b=3$, density-biased, ($\Gamma$, $\Omega_0$, $\lambda_0$) = 
(0.25, 1, 0) model.  Although the model tends to overproduce voids at
the smaller values of $\bar d$, it predicts large sample-to-sample variations
in the VPF (wide error corridors), so it cannot be ruled out by the
current data.

}
\vfill
\endinsert
\textskip

\topinsert
\capskip
\centerline{
\epsfysize=6.8truein
\epsfbox[0 0 612 774]{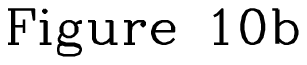}
}
{\eightpoint
\capskip
\noindent {\bf Figure 10} --- {\it continued}
}
\endinsert
\textskip

\topinsert
\capskip
\centerline{
\epsfysize=6.8truein
\epsfbox[0 0 612 774]{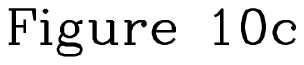}
}
{\eightpoint
\capskip
\noindent {\bf Figure 10} --- {\it continued}
}
\endinsert
\textskip

Beginning with Figure 10a, we see that the VGH data generally lie between
the 75\% and 95\% VPFs of 
the unbiased, ($\Gamma$, $\Omega_0$, $\lambda_0$) = (0.25, 0.3, 0.7) model.
Although the model systematically underproduces large voids on average,
it is consistent with the data at the $2\sigma$ level.
The only exception is the large-$R$ tail of the $\bar d=4.5$ $\hmpc$, northern
VPF, which lies slightly above the model's 95\% VPF.  This tail is 
the one region of qualitative discrepancy between the VPFs of the CfA2
North sample and the corresponding VPFs of the other two VGH samples, so
we are reluctant to place much weight on it.  Putting it aside, we conclude
that the VGH data are consistent with an unbiased Gaussian 
model, and that the voids in the CfA2 survey provide no convincing
evidence for biased galaxy formation.

Figure 10b shows still better agreement between model and data.  Most of 
the VGH results lie between the 25\% and 75\% VPFs of this $b=1.5$,
C/O-biased, standard CDM model.  Again the tail of the $\bar d=4.5$ $\hmpc$,
northern VPF lies just above the model's 95\% VPF.  Apart from this 
discrepancy, the VGH data agree well with this Gaussian model, which
incorporates the sort of modest bias predicted by cosmological simulations
with gas dynamics (CO; Katz \etal 1992).
Similar results obtain for peaks-biased models.

Figure 10c shows results for our most severely biased \gam\ model.
For $\bar d \leq 7.4$ $\hmpc$, the VGH data lie mostly between the 
5\% and 25\% VPFs, confirming the tendency of this model to overproduce
large voids.  However, even though the mean VPF predicted by this model
is a poor match to the data (see Figure 9b), the model remains consistent
with the current observations because it predicts large VPF variations 
in samples of this size.  This is probably a specific example of 
a general problem: models with large voids also have large sample-to-sample
fluctuations in the VPF, so larger data sets are needed to rule them out.

The overall message of this section may seem somewhat discouraging.  The data
do not convincingly exclude any of these three models, even though the
models span a large range in predicted VPFs.
However, the situation should improve in the near future, as the CfA 
extension has now been completed over a larger fraction of the sky.
M.\ Vogeley (private communication) reports that VGH have now doubled
the size of their observational sample, and that the VPF results for
this larger sample are similar to those for the original data set.
With the doubled sample, it may well be possible to rule out severely biased
models, and perhaps even unbiased models.
In the longer run, massive redshift surveys like the Sloan Digital Sky Survey
(Gunn \& Knapp 1993)
will probe much greater volumes and yield highly accurate estimates of
the VPF, placing correspondingly tight constraints on theories of
biased galaxy formation.

\subsectionbegin {3.6 The Probability Function of Underdense Regions} 

Figure 11 illustrates
the `underdense probability function' $P_{80}(R)$ for a variety of our 
\gam\ models.  Its three rows display
redshift-space UPFs for 
our
$(\Gamma$, $\Omega_0$, $\lambda_0$) = 
(0.5, 1, 0),
(0.25, 1, 0), and
(0.25, 0.3, 0.7) models, from top to bottom.
The figure's columns and
line types have the same biasing associations as 
Figure 9's (including the association of the dot-dashed line in the 
top right panel with $b=1.5$, C/O biasing).
Since it is always
easier to find an $80\%$ underdense region than a totally empty one,
the UPFs 
have higher amplitudes at a given radius than their
corresponding VPFs (cf. Figure 9). 
Also, these redshift-space 
UPFs 
-- like redshift-space VPFs --
are somewhat
larger than their real-space counterparts (cf. Figure 3).

\topinsert
\capskip
\centerline{
\epsfxsize=4.9truein
\epsfbox[0 180 612 774]{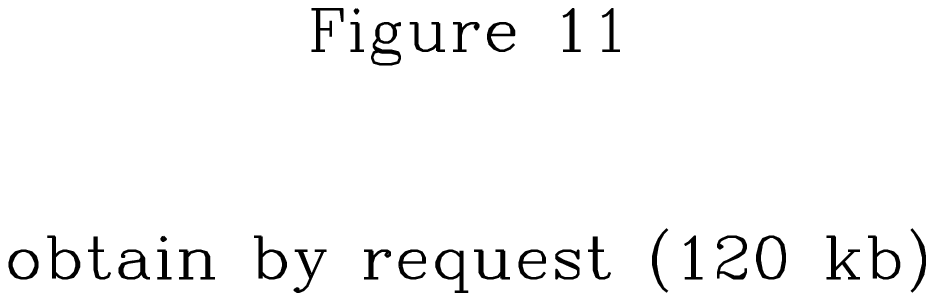}
}
{\eightpoint
\capskip
\noindent {\bf Figure 11} --- The underdense probability functions (UPFs)
of our \gam\ models.
$P_{80}(R)$ is the probability that the average galaxy density within a
randomly placed sphere of radius $R$ (in $h^{-1}$ Mpc) is more than 80\%
below the global mean density. Unlike the VPF, the UPF is independent of
the space density of the tracer population, except for shot noise. 
In each panel,
solid, dotted, short-dashed, and long-dashed lines represent UPFs for
models with $b=1$, 1.5, 2, and 3, respectively. The $b\geq 1.5$
lines in the left and right columns are associated with peaks biasing
and density biasing, respectively. The dot-dashed line in the upper
right panel represents the $b=1.5$, C/O-biased model.
The model parameters $\Gamma$, $\Omega_0$, and $\lambda_0$ are
listed in each panel. 
}
\endinsert
\textskip

Figure 11 confirms several of the $\overline
d$-independent trends of Figure 9. For example:
({\it i\/}) 
density biasing produces significantly larger voids than
peaks biasing;
({\it ii\/}) 
peaks biasing and C/O biasing produce
statistically similar
voids;
({\it iii\/}) among biased models,
the sizes of voids generally
increase as $\Gamma$ is lowered;
({\it iv\/}) the 
differences in void structure
between $b=1$ and $b=1.5$ models can be much larger than those
among $b\geq 1.5$ models, and this effect is stronger among
models with more large-scale power.
The UPF of unbiased models increases with increasing large-scale power,
\ie it is higher for lower $\Gamma$.  WC found a similar result for 
power-law models.  This trend is opposite to that of the unbiased VPF
(again consistent with WC).

One advantage of the UPF over the VPF is its simplicity.  It is independent
of galaxy density, except for shot noise, which becomes progressively less
important at larger radii.  As a result, a single UPF can be measured
directly from magnitude-limited data, without defining multiple volume-limited
samples that have multiple values of $\bar d$.
(However, additional complications enter if galaxy clustering depends
systematically on luminosity.)
Because the UPF falls more slowly with radius than the VPF, it can be
measured out to larger scales in a given observational sample.
We do not know of any existing observational data on the UPF, so the 
curves in Figure 11 will have to stand as blind predictions.
UPF results from the CfA2 survey should be available in the near future
(M.\ Vogeley, private communication).

\sectionbegin {\bf 4. Conclusions} 

We have investigated the behaviour of the void probability function (VPF) in
a wide range of initially Gaussian models for the origin of
large-scale structure.  We have focused on the sensitivity of the VPF
to assumptions about biased galaxy formation.  Our main conclusions
about the dependence of the VPF on model parameters are as follows.

\item {1.} Peculiar velocity distortions have relatively little effect on
the VPF, though VPFs measured in redshift space tend to be slightly
higher than those measured in real space.

\item {2.} The real-space VPF is extremely
insensitive to the cosmic density parameter $\Omega_0$ and the 
cosmological constant $\lambda_0$, provided that models with different
values of these cosmological parameters have the same
initial power spectrum and biasing prescription and are normalized
to the same rms mass fluctuation at the present epoch.

\item {3.} In the absence of biasing, the VPF depends only weakly on
the shape of the initial power spectrum, provided that models are
normalized to $\sigma_{8,{\rm mass}} \approx 1$ today.  Unbiased models
with more small-scale power have slightly higher VPFs.

\item {4.} The VPF is quite sensitive to the prescription adopted for
biased galaxy formation.  Density-biased models produce much larger voids
than peaks-biased models, or models that incorporate CO's non-linear
biasing prescription.  Peaks-biased models have higher VPFs
than unbiased models.

\item {5.} The VPF depends more strongly on the form of the biasing
prescription (\eg density biasing versus peaks biasing) than on the
bias factor $b$; for a given biasing scheme, the VPF does not change
much in the range $1.5 \leq b \leq 3$.  Given the physical uncertainty
in the appropriate choice of biasing scheme, the VPF is probably not
a useful tool for determining $b$.  However, the VPF can distinguish
unbiased models from some biased models, and it can place useful constraints
on the relation between galaxies and mass.

\item {6.} Conclusions $1-5$ also apply to the underdense region probability
function (UPF), except that the trend of the unbiased VPF with power spectrum
(conclusion 3) is reversed for the UPF.

We have compared our model predictions to the data of VGH, who have measured
VPFs of the original CfA survey and the completed regions of the
CfA extension.  Our main conclusions from comparison between models and
the VGH data are as follows.

\item {7.} Unbiased Gaussian models reproduce the observed VPFs fairly well.
The void probability data from the CfA survey do not provide compelling
evidence for biased galaxy formation.

\item {8.} Models that incorporate moderate levels of biasing,
similar to those predicted by cosmological simulations with gas dynamics,
produce the best overall fit to the VGH data.  

\item {9.} Density-biased models tend to overproduce large voids, at least
on average.

\item {10.} Large finite volume fluctuations are expected in samples of the
size analyzed by VGH, and models that predict the largest voids also 
predict the largest sample-to-sample variations in the VPF.
As a result, the VGH data do not convincingly rule out any of our models.

The opening paragraph of this paper posed two broad questions.
Can the gravitational growth of Gaussian primordial fluctuations account for 
the observed voids, or do initially Gaussian models require that galaxy 
formation be
suppressed in low density regions in order to produce voids as large and
empty as observed?  Do voids represent regions where there is no
mass, or merely regions where there are no (bright) galaxies?  
Conclusion (10) tells us that we cannot offer convincing answers to these
questions with the present observational data, at least not using the
VPF as our primary tool.  The VGH data are consistent with the hypothesis
that voids in the CfA survey grew gravitationally from Gaussian primordial 
fluctuations, and that these voids are truly as underdense in mass as they
are in galaxies.  However, the data also permit a substantial bias between
galaxies and mass, with the voids underdense, but not nearly so empty
as they appear.  Analysis of future redshift surveys will help 
to distinguish these possibilities.

\sectionbegin{Acknowledgments}

We are grateful to Michael Vogeley for helpful discussions about
the CfA results and for 
providing us with the VGH data in convenient form.  We are also grateful
to Changbom Park for allowing us to use his PM code, and to Renyue
Cen for computing the parameters of equation (1) for our cubic cell filter.
BL acknowledges the support of the Institute of Astronomy in Cambridge 
and a SERC postdoctoral fellowship.
DHW is a W.M.\ Keck Foundation fellow at the Institute for Advanced Study.
He acknowledges additional support from the Ambrose Monell Foundation 
and from the U.S.\ National Science Foundation in the form of grant
PHY92-45317 and a NATO postdoctoral fellowship to Cambridge University,
where this work began.

\vfill\eject

\sectionbegin {\bf References} 

\refbook {Bardeen, J.~M., Bond, J.~R., Kaiser, N. \& Szalay, A.~S.,
1986. {\it Astrophys.
J.\/}, {\bf 304}, 15 (BBKS). } 

\refbook {Betancort-Rijo, J., 1990.
{\it Mon. Not. R. astr. Soc.\/}, {\bf 246}, 608. } 

\refbook {Bhavsar, S.~P., Gott, J.~R. \& Aarseth, S.~J., 1981. {\it Astrophys.
J.\/}, {\bf 246}, 656. } 

\refbook {Bond, J.~R. \& Efstathiou, G., 1991. {\it Phys. Lett. B\/}, {\bf 
265}, 245. }

\refbook {Bouchet, F., Strauss, M.~A., Davis, M., Fisher, K.~B., Yahil, A. 
\& Huchra, J.~P., 1993. submitted to {\it Astrophys. J.\/} }

\refbook {Cen, R. \& Ostriker, J.~P., 1992. {\it Astrophys. J.
Lett.\/}, {\bf 399}, L113. } 

\refbook {Cen, R. \& Ostriker, J.~P., 1993.
Princeton Observatory Preprint POP-493, submitted to {\it Astrophys. J.\/} 
(CO).}

\refbook {Cole, S. \& Kaiser, N., 1989.
{\it Mon. Not. R. astr. Soc.\/}, {\bf 237}, 1127. } 

\refbook {Davis, M., Efstathiou, G., Frenk, C.~S. \& White, S.~D.~M., 
1985. {\it Astrophys. J.\/}, {\bf 292}, 371 (DEFW). }

\refbook {Davis, M., Huchra, J., Latham, D.~W. \& 
Tonry, J., 1982. {\it Astrophys. J.\/},
{\bf 253}, 423. } 

\refbook {Davis, M. \& Peebles, P.~J.~E., 1983. {\it Astrophys. J.\/},
{\bf 267}, 465. } 

\refbook {Davis, M., Summers, F.~J. \& Schlegel, D., 1992.
{\it Nature\/}, {\bf 359}, 393. }

\refbook {Davis, R.L., Hodges, H.~M., Smoot, G.~F.,
Steinhardt, P.~J. \& Turner, M.~S., 1992.
{\it Phys. Rev. Lett.\/}, {\bf 69}, 1856. }

\refbook {Dekel, A. \& Rees, M.~J., 1987.
{\it Nature\/}, {\bf 326}, 455. }

\refbook {Dekel, A. \& Silk, J., 1986.
{\it Astrophys. J.\/}, {\bf 303}, 39. }

\refbook {de Lapparent, V., Geller, M.~J. \& Huchra, J.~P., 1986.
{\it Astrophys. J. Lett.\/}, {\bf 302}, L1. }

\refbook {de Lapparent, V., Geller, M.~J. \& Huchra, J.~P., 1991.
{\it Astrophys. J.\/}, {\bf 369}, 273. }

\refbook {Doroshkevich, A.~G., Kotok, E.~V., Novikov, I.~D., Polyudov, 
A.~N., Shandarin, S.~F., \& Sigov, Yu.~S., 1980.
{\it Mon. Not. R. astr. Soc.\/}, {\bf 192}, 321. } 

\refbook {Dressler, A., 1980. {\it Astrophys. J.\/}, {\bf 236}, 351. }

\refbook {Dubinski, J., 
da Costa, N., Goldwirth, D.~S., Lecar, M. \& Piran, T., 1993.
{\it Astrophys. J.\/}, in press. }

\refbook {Efstathiou, G., Bond, J.~R. \& White, S.~D.~M.,
1992.
{\it Mon. Not. R. astr. Soc.\/}, {\bf 258}, 1P (EBW). } 

\refbook {Efstathiou, G., Davis, M., Frenk, C.~S. \& White, S.~D.~M.,
1985. {\it Astrophys. J. Supp.\/}, {\bf 57}, 241. } 

\refbook {Efstathiou, G., Kaiser, N., Saunders, W., Lawrence, A., 
Rowan-Robinson, M., Ellis, R.~S. \& Frenk, C.~S., 1990.
{\it Mon. Not. R. astr. Soc.\/}, {\bf 247}, 10P. } 

\refbook {Einasto, J., Einasto, M., Gramann, M. \& Saar E., 1991.
{\it Mon. Not. R. astr. Soc.\/}, {\bf 248}, 593 (EEGS). } 

\refbook {Evrard, A.~E., Summers, F.~J. \& Davis, M., 1993. 
submitted to {\it Astrophys. J.\/}.}

\refbook {Fisher, K.~B., Davis, M., Strauss, M.~A., Yahil, A. \& 
Huchra, J.~P., 1993.
{\it Astrophys. J.\/}, {\bf 402}, 42. }

\refbook {Fry, J.~N., 1986.
{\it Astrophys. J.\/}, {\bf 306}, 358. }

\refbook {Fry, J.~N., Giovanelli, R., Haynes, M.~P., Melott, A.~L.,
\& Scherrer, R.~J., 1989.
{\it Astrophys. J.\/}, {\bf 340}, 11. }

\refbook {Gelb, J.~M., 1992. {\it Ph.D. thesis\/}, MIT. } 

\refbook {Geller, M.~J. \& Huchra, J.~P., 1989.
{\it Science\/}, {\bf 246}, 897. }

\refbook {Gott, J.~R., Mao, S., Park, C., \& Lahav, O., 1992.
{\it Astrophys. J.\/}, {\bf 385}, 26. }

\refbook {Gott, J.~R., Melott, A.~L. \& Dickinson, M., 1986.
{\it Astrophys. J.\/}, {\bf 306}, 341. }

\refbook {Gott, J.~R. {\it et al.\/}, 1989. 
{\it Astrophys. J.\/}, {\bf 340}, 625. } 

\refbook {Gott, J.~R. \& Rees, M.~J., 1975. {\it Astr. Astrophys.\/},
{\bf 45}, 365. } 

\refbook {Gott, J.~R. \& Turner, E.~L., 1977. {\it Astrophys. J.\/}, {\bf 216},
357. } 

\refbook {Gregory, S.~A. \& Thompson, L.~A., 1978. 
{\it Astrophys. J.\/}, {\bf 222},
784. } 

\refbook {Gunn, J.~E., 1982. In: {\it Astrophysical 
Cosmology\/}, p.233, eds Br\"uck, H.~A., Coyne, G.~V. \& Longair, 
M.~S.,
Specola Vaticana, Rome.} 

\refbook {Gunn, J.~E. \& Knapp, G.~R., 1993. In: {\it Astronomical
Surveys\/}, ed Soifer, B.~T., ASP conference proceedings, in press.}

\refbook {Haynes, M.~P. \& Giovanelli, R., 1986.
{\it Astrophys. J. Lett.\/}, {\bf 306}, L55. }

\refbook {Haynes, M.~P. \& Giovanelli, R., 1988. In:
{\it Large-Scale Motions in the Universe\/},
p. 31, eds
Rubin, V.~C. \& Coyne, G.~V., Princeton University Press, Princeton. }

\refbook {Hernquist, L., Bouchet, F.~R. \& Suto, Y., 1991.
{\it Astrophys. J. Supp.\/}, {\bf 75}, 231. } 

\refbook {Huchra, J.~P., Davis, M., Latham, D. \& Tonry, J., 1983.
{\it Astrophys. J. Supp.\/}, {\bf 52}, 89. } 

\refbook {Joeveer, M. \& Einasto, J., 1978. In:
{\it The Large Scale Structure of the Universe,
IAU Symposium 79\/}, p. 241, eds
Longair, M.~S. \& Einasto, J., Reidel, Dordrecht. }

\refbook {Kaiser, N., 1984. 
{\it Astrophys. J. Lett.\/}, {\bf 284}, L9. }

\refbook {Kaiser, N., 1988. In:
{\it Evolution of Large Scale Structures in the Universe,
IAU Symposium 130\/}, p. 43, eds Audouze, J. \& Szalay, A.,
Reidel, Dordrecht. }

\refbook {Katz, N., Hernquist, L. \& Weinberg, D.~H., 1992. 
{\it Astrophys. J. Lett.\/}, {\bf 399}, L109. }

\refbook {Katz, N., Quinn, T. \& Gelb, J.~M., 1993. 
submitted to {\it Mon. Not. R. astr. Soc.\/}.}

\refbook {Kauffmann, G. \& Melott, A.~L., 1992. 
{\it Astrophys. J.\/}, {\bf 393}, 415. }

\refbook {Kirshner, R.~P., Oemler, A., Schechter, P.~L. \& Shectman, 
S.~A., 1981.
{\it Astrophys. J. Lett.\/}, {\bf 248}, L57. }

\refbook {Kirshner, R.~P., Oemler, A., Schechter, P.~L. \& Shectman, 
S.~A., 1987.
{\it Astrophys. J.\/}, {\bf 314}, 493. }

\refbook {Klypin, A., Holtzman, J., Primack, J. \& Reg\"os, E., 1993.
submitted to {\it Astrophys. J.\/}.}

\refbook {Lachi\`eze-Rey, M., da Costa, L.~N. \& Maurogordato, S., 1992.
{\it Astrophys. J.\/}, {\bf 399}, 10. }

\refbook {Liddle, A.~R. \& Lyth, D.~H., 1992.
{\it Phys. Lett. B\/}, {\bf 291}, 391. }

\refbook {Lidsey, J.~E. \& Coles, P., 1992.
{\it Mon. Not. R. astr. Soc.\/}, {\bf 258}, 57P. } 

\refbook {Lucchin, F., Mataresse, S. \& Mollerach, S., 1992.
{\it Astrophys. J. Lett.\/}, {\bf 401}, L49. }

\refbook {Maddox, S.~J., Efstathiou, G., Sutherland, W.~J. \& Loveday, 
J., 1990.
{\it Mon. Not. R. astr. Soc.\/}, {\bf 242}, 43P. } 

\refbook {Maurogordato, S. \&
Lachi\`eze-Rey, M., 1987.
{\it Astrophys. J.\/}, {\bf 320}, 13.}

\refbook {Melott, A.~L., 1986. 
{\it Phys. Rev. Lett.\/}, {\bf 56}, 1992. } 

\refbook {Moore, B., Frenk, C.~S., Weinberg, D.~H., Saunders, W.,
Lawrence, A., Ellis, R.~S., Kaiser, N., Efstathiou, G. \& 
Rowan-Robinson, M., 1992.
{\it Mon. Not. R. astr. Soc.\/}, {\bf 256}, 477. } 

\refbook {Narlikar, J.~V. \& Padmanabhan, T., 1991.
{\it Annu. Rev. Astron. Astrophys.\/}, {\bf 29}, 325. }

\refbook {Park, C., 1990. {\it Ph.D. thesis\/}, Princeton University. } 

\refbook {Park, C., 1991.
{\it Mon. Not. R. astr. Soc.\/}, {\bf 251}, 167. } 

\refbook {Postman, M. \& Geller, M.J., 1984. 
{\it Astrophys. J.\/}, {\bf 281}, 95. }

\refbook {Rees, M.~J., 1985.
{\it Mon. Not. R. astr. Soc.\/}, {\bf 213}, 75P. } 

\refbook {Reg\"os, E. \& Geller, M.~J., 1991.
{\it Astrophys. J.\/}, {\bf 377}, 14. }

\refbook {Rood, H.~J., 1988.
{\it Annu. Rev. Astron. Astrophys.\/}, {\bf 26}, 245. }

\refbook {Salopek, D.~S., 1992.
{\it Phys. Rev. Lett.\/}, {\bf 69}, 3602. }

\refbook {Saunders, W., Frenk, C., Rowan-Robinson, M., Efstathiou, 
G., Lawrence, A., Kaiser, N., Ellis, R., Crawford, J.,
Xia, X.~Y. \& Parry, I., 1991.
{\it Nature\/}, {\bf 349}, 32. }

\refbook {Silk, J., 1985.
{\it Astrophys. J.\/}, {\bf 297}, 1. }

\refbook {Smoot, G.~F. {\it et al.\/}, 1992.
{\it Astrophys. J. Lett.\/}, {\bf 396}, L1. }

\refbook {Souradeep, T. \& Sahni, V., 1992.
{\it Mod. Phys. Lett. A\/}, {\bf 7}, 3541. }

\refbook {Taylor, A.~N. \& Rowan-Robinson, M., 1992.
{\it Nature\/}, {\bf 359}, 396. }

\refbook {van de Weygaert, R. \& van Kampen, E., 1993. 
{\it Astrophys. J.\/}, in press. } 

\refbook {Vogeley, M.~S., Geller, M.~J. \& Huchra, J.~P., 1991.
{\it Astrophys. J.\/}, {\bf 382}, 44 (VGH). } 

\refbook {Vogeley, M.~S., Park, C., Geller, M.~J. \& Huchra, J.~P., 1992.
{\it Astrophys. J. Lett.\/}, {\bf 391}, L5. }

\refbook {Weinberg, D.~H. \& Cole, S., 1992.
{\it Mon. Not. R. astr. Soc.\/}, {\bf 259}, 652 (WC). } 

\refbook {Weinberg, D.~H. \& Gunn, J.~E., 1990. 
{\it Mon. Not. R. astr. Soc.\/}, {\bf 247}, 260. } 

\refbook {West, M.~J., Weinberg, D.~H. \& Dekel, A., 1990. 
{\it Astrophys. J.\/}, {\bf 353}, 329. } 

\refbook {White, S.~D.~M., 1979.
{\it Mon. Not. R. astr. Soc.\/}, {\bf 186}, 145. } 

\refbook {White, S.~D.~M., Davis, M., Efstathiou, G. \& Frenk, C.~S., 1987. 
{\it Nature\/}, {\bf 330}, 451. }

\bye